\documentclass[]{aastex631}

\usepackage{amsmath}
\usepackage{todonotes}
\usepackage{comment}

\begin{document}

\title{Debiasing the Observed Fast Radio Burst Population with the CHIME/FRB Selection Function}

\correspondingauthor{Kyle McGregor}
\email{kyle.mcgregor@mail.mcgill.ca}

\author[0000-0003-2111-3437]{Kyle McGregor}
\affiliation{Trottier Space Institute at McGill University, 3550 rue University, Montr\'eal, QC H3A 2A7, Canada}
\affiliation{Department of Physics, McGill University, 3600 rue University, Montr\'eal, QC H3A 2T8, Canada}

\author[0000-0003-2317-1446]{Jason W. T. Hessels}
\affiliation{Trottier Space Institute at McGill University, 3550 rue University, Montr\'eal, QC H3A 2A7, Canada}
\affiliation{Department of Physics, McGill University, 3600 rue University, Montr\'eal, QC H3A 2T8, Canada}
\affiliation{Anton Pannekoek Institute for Astronomy, University of Amsterdam, Science Park 904, 1098 XH, Amsterdam, The Netherlands}
\affiliation{ASTRON, Netherlands Institute for Radio Astronomy, Oude Hoogeveensedijk 4, 7991 PD Dwingeloo, The Netherlands}

\author[0000-0001-9345-0307]{Victoria M. Kaspi}
\affiliation{Trottier Space Institute at McGill University, 3550 rue University, Montr\'eal, QC H3A 2A7, Canada}
\affiliation{Department of Physics, McGill University, 3600 rue University, Montr\'eal, QC H3A 2T8, Canada}

\author[0000-0002-6823-2073]{Kaitlyn Shin}
\affiliation{Cahill Center for Astronomy and Astrophysics, MC 249-17 California Institute of Technology, Pasadena CA 91125, USA}

\author[0000-0003-4098-5222]{Fengqiu Adam Dong}
\affiliation{Department of Physics and Astronomy, York University, 4700 Keele Street, Toronto, Ontario, ON MJ3 1P3, Canada}

\author[0009-0009-0938-1595]{Naman Jain}
\affiliation{Trottier Space Institute at McGill University, 3550 rue University, Montr\'eal, QC H3A 2A7, Canada}
\affiliation{Department of Physics, McGill University, 3600 rue University, Montr\'eal, QC H3A 2T8, Canada}

\author[0000-0002-7164-9507]{Robert Main}
\affiliation{Trottier Space Institute at McGill University, 3550 rue University, Montr\'eal, QC H3A 2A7, Canada}
\affiliation{Department of Physics, McGill University, 3600 rue University, Montr\'eal, QC H3A 2T8, Canada}

\author[0000-0002-4623-5329]{Mawson Sammons}
\affiliation{Trottier Space Institute at McGill University, 3550 rue University, Montr\'eal, QC H3A 2A7, Canada}
\affiliation{Department of Physics, McGill University, 3600 rue University, Montr\'eal, QC H3A 2T8, Canada}

\author[0009-0001-3334-9482]{Michele Woodland}
\affiliation{Department of Astronomy \& Astrophysics, University of California, Santa Cruz, 1156 High Street, Santa Cruz, CA 95064, USA}

\author[0009-0005-5370-7653]{Daniel Amouyal}
\affiliation{Trottier Space Institute at McGill University, 3550 rue University, Montr\'eal, QC H3A 2A7, Canada}
\affiliation{Department of Physics, McGill University, 3600 rue University, Montr\'eal, QC H3A 2T8, Canada}

\author[0000-0001-5628-7256]{Derek Bingham}
\affiliation{Department of Statistics and Actuarial Science, Simon Fraser University, 8888 University Dr W, Burnaby, BC V5A 1S6, Canada}

\author[0000-0002-1800-8233]{Charanjot Brar}
\affiliation{NRC Herzberg Astronomy and Astrophysics, 5071 West Saanich Road, Victoria, BC V9E2E7, Canada}

\author[0000-0001-6422-8125]{Amanda M. Cook}
\affiliation{Department of Physics, McGill University, 3600 rue University, Montr\'eal, QC H3A 2T8, Canada}
\affiliation{Trottier Space Institute at McGill University, 3550 rue University, Montr\'eal, QC H3A 2A7, Canada}
\affiliation{Anton Pannekoek Institute for Astronomy, University of Amsterdam, Science Park 904, 1098 XH, Amsterdam, The Netherlands}

\author[0000-0002-1348-8063]{Radu V. Craiu}
\affiliation{Department of Statistical Sciences, University of Toronto, 700 University Ave, Toronto ON M4G 1Z5, Canada}

\author[0000-0002-8376-1563]{Alice Curtin}
\affiliation{Trottier Space Institute at McGill University, 3550 rue University, Montr\'eal, QC H3A 2A7, Canada}
\affiliation{Department of Physics, McGill University, 3600 rue University, Montr\'eal, QC H3A 2T8, Canada}
\affiliation{Anton Pannekoek Institute for Astronomy, University of Amsterdam, Science Park 904, 1098 XH, Amsterdam, The Netherlands}

\author[0000-0003-3734-8177]{Gwendolyn Eadie}
\affiliation{David A.\ Dunlap Institute Department of Astronomy \& Astrophysics, University of Toronto, 50 St. George Street, Toronto, ON M5S 3H4, Canada}
\affiliation{Department of Statistical Sciences, University of Toronto, 700 University Ave, Toronto ON M4G 1Z5, Canada}

\author[0000-0002-3382-9558]{B.~M. Gaensler}
\affiliation{Department of Astronomy \& Astrophysics, University of California, Santa Cruz, 1156 High Street, Santa Cruz, CA 95064, USA}
\affiliation{Dunlap Institute for Astronomy \& Astrophysics, University of Toronto, 50 St.~George Street, Toronto, ON M5S 3H4, Canada}
\affiliation{David A.\ Dunlap Institute Department of Astronomy \& Astrophysics, University of Toronto, 50 St. George Street, Toronto, ON M5S 3H4, Canada}

\author[0000-0002-8043-0048]{Jeff Huang}
\affiliation{Trottier Space Institute at McGill University, 3550 rue University, Montr\'eal, QC H3A 2A7, Canada}
\affiliation{Department of Physics, McGill University, 3600 rue University, Montr\'eal, QC H3A 2T8, Canada}

\author[0009-0004-4176-0062]{Afrokk Khan}
\affiliation{Trottier Space Institute at McGill University, 3550 rue University, Montr\'eal, QC H3A 2A7, Canada}
\affiliation{Department of Physics, McGill University, 3600 rue University, Montr\'eal, QC H3A 2T8, Canada}

\author[0000-0002-4209-7408]{Calvin Leung}
\affiliation{Department of Astronomy, University of California, Berkeley, CA 94720, United States}
\affiliation{Miller Institute for Basic Research, University of California, Berkeley, CA 94720, United States}

\author[0000-0002-4279-6946]{Kiyoshi W. Masui}
\affiliation{MIT Kavli Institute for Astrophysics and Space Research, Massachusetts Institute of Technology, 77 Massachusetts Ave, Cambridge, MA 02139, USA}
\affiliation{Department of Physics, Massachusetts Institute of Technology, 77 Massachusetts Ave, Cambridge, MA 02139, USA}

\author[0000-0002-8897-1973]{Ayush Pandhi}
\affiliation{Trottier Space Institute at McGill University, 3550 rue University, Montr\'eal, QC H3A 2A7, Canada}
\affiliation{Department of Physics, McGill University, 3600 rue University, Montr\'eal, QC H3A 2T8, Canada}

\author[0009-0008-7264-1778]{Swarali S. Patil}
\affiliation{Department of Physics and Astronomy, West Virginia University, PO Box 6315, Morgantown, WV 26506, USA}
\affiliation{Center for Gravitational Waves and Cosmology, West Virginia University, Chestnut Ridge Research Building, Morgantown, WV 26505, USA}

\author[0000-0002-8912-0732]{Aaron B. Pearlman}
\altaffiliation{NASA Hubble Fellow}
\affiliation{MIT Kavli Institute for Astrophysics and Space Research, Massachusetts Institute of Technology, 77 Massachusetts Ave, Cambridge, MA 02139, USA}
\affiliation{Department of Physics, Massachusetts Institute of Technology, 77 Massachusetts Ave, Cambridge, MA 02139, USA}
\affiliation{Department of Physics, McGill University, 3600 rue University, Montr\'eal, QC H3A 2T8, Canada}
\affiliation{Trottier Space Institute at McGill University, 3550 rue University, Montr\'eal, QC H3A 2A7, Canada}

\author[0009-0008-2000-6959]{Sachin Pradeep E.~T.}
\affiliation{Trottier Space Institute at McGill University, 3550 rue University, Montr\'eal, QC H3A 2A7, Canada}
\affiliation{Department of Physics, McGill University, 3600 rue University, Montr\'eal, QC H3A 2T8, Canada}

\author[0000-0002-7374-7119]{Paul Scholz}
\affiliation{Department of Physics and Astronomy, York University, 4700 Keele Street, Toronto, Ontario, ON MJ3 1P3, Canada}

\author[0000-0003-2631-6217]{Seth Siegel}
\affiliation{SKA Observatory, Science Operations Centre, CSIRO ARRC, 26 Dick Perry Avenue, Kensington, WA 6151, Australia}
\affiliation{Perimeter Institute for Theoretical Physics, 31 Caroline Street N, Waterloo, ON N25 2YL, Canada}
\affiliation{Trottier Space Institute at McGill University, 3550 rue University, Montr\'eal, QC H3A 2A7, Canada}
\affiliation{Department of Physics, McGill University, 3600 rue University, Montr\'eal, QC H3A 2T8, Canada}

\author[0000-0002-2088-3125]{Kendrick M. Smith}
\affiliation{Perimeter Institute for Theoretical Physics, 31 Caroline Street N, Waterloo, ON N25 2YL, Canada}

\author[0000-0002-9761-4353]{David Stenning}
\affiliation{Department of Statistics and Actuarial Science, Simon Fraser University, 8888 University Dr W, Burnaby, BC V5A 1S6, Canada}



\begin{abstract}
The recent release of CHIME/FRB Catalog~2 provides the largest sample to date with which to investigate the intrinsic distributions of fast radio bursts (FRBs). Leveraging an expanded campaign of 587,367 synethetic bursts injected into the live CHIME/FRB search pipeline, we perform a population analysis of the fluence, scattering timescale, pulse width, and dispersion measure distributions of Catalog~2 FRBs. We first infer the intrinsic population using a resampling-based framework that accounts for instrumental selection effects following previous CHIME/FRB population studies. A central goal of this work is to constrain the intrinsic distribution of scattering timescales, that remained weakly constrained in Catalog~1 owing to limited statistics at moderate and large scattering times ($\tau \gtrsim 10\,\mathrm{ms}$ at 600~MHz) and sparse injection coverage in this regime. Second, we construct an explicit multidimensional selection function by training a logistic regression model on the injected events. This model estimates the detection probability as a function of FRB observable properties, including higher-order interaction terms. We incorporate this selection function into a simulation-based inference framework to refine the inferred intrinsic scattering-timescale distribution. We find evidence for a slight downturn in the intrinsic FRB scattering timescale distribution, though a flat or slightly rising distribution cannot be ruled out, that is further supported through a comparison with the higher-frequency scattering timescale distribution observed by Commensal Real-time ASKAP Fast Transients (CRAFT) survey.
\end{abstract}

\keywords{Radio Transient Sources (2008)}


\section{Introduction} \label{sec:intro}

Fast radio bursts (FRBs) are bright, millisecond-scale radio transients that have emerged as an intriguing frontier of time-domain survey astrophysics. First discovered nearly 20 years ago \citep{Lorimer2007}, the study of FRBs has matured from a handful of enigmatic discoveries into a statistically rich population with the advent of wide-field radio surveys. In particular, with the release of more than 4500 FRBs observed by the Canadian Hydrogen Intensity Mapping Experiment \citep[hereafter referred to as CHIME/FRB Catalog 2;][]{Catalog2} the cumulative number of known FRB sources has greatly increased, enabling population-level analyses with improved statistical precision relative to earlier studies \citep[][Jain et al. in prep.]{2025arXiv250608932W, 2026ApJ...997L...5P, Catalog2, 2026arXiv260219335S}. 

Selection effects are inherent to every astrophysical survey and are implicitly encoded in the sample detected by CHIME/FRB. Correcting for these effects is therefore imperative for making unbiased inferences about the underlying FRB population. Formally, these corrections are encoded through a survey’s selection function, which specifies the probability that an instrument will detect a given source with certain incident properties. 
With the emergence of large-scale, real-time transient search pipelines \citep[e.g.][]{CHIMEFRBSystemOverview, 2023MNRAS.524.4275J, 2024ApJ...967...29L, 2025PASA...42...36S, 2026MNRAS.548ag665K}, the characterization of selection effects has become increasingly complex, as detection is mediated by multi-stage triggering, hierarchical data products, and nonlinear interactions between burst properties and search algorithms. Complex radio-frequency interference environments may also cause false positive and false negative triggers  \citep[e.g.][]{2015MNRAS.451.3933P, 2026AJ....171...73K, RBFLOAT}. This motivates the development of empirical selection function modeling, which in FRB and pulsar surveys is routinely achieved through injection-based tests of search pipelines \citep[e.g.,][]{2015ApJ...812...81L, 2021ApJS..257...59C, 2023MNRAS.523.5109Q}. For these studies, synthetic signals are injected directly into the data streams and processed through the search pipeline to empirically measure detection efficiency.

To correct for selection biases in the first large-scale catalog of CHIME/FRB sources \citep{2021ApJS..257...59C} ($\sim\!500$ FRBs, hereafter referred to as Catalog~1), a population synthesis was performed using a large dataset of synthetic bursts injected into the instrument’s real-time live search pipeline (hereafter, {\it injections}). This dataset, along with the observed catalog, was applied in Section~6 of \citet{2021ApJS..257...59C} to infer a fiducial, selection-corrected model for the marginalized distributions of FRB observables; the injections were subsequently leveraged by \citet{ShinEtAl} to constrain the redshift and energy distributions of FRBs. The design and implementation of CHIME/FRB's injection pipeline, as well as a detailed description of the resulting injection dataset and its intended use, were presented by \citet{Merryfield}. In the work here, we build on the \citet{2021ApJS..257...59C} analysis by fitting a parametric selection function model that explicitly accounts for correlations between observables, employing a logistic regression framework to more accurately capture the structure of the CHIME/FRB detection process and extend the range of the parameter space to better quantify CHIME/FRB's sensitivity to highly-scattered FRBs. This is in accompaniment to fitting an analogous single-variate model to \citet{2021ApJS..257...59C} using Catalog~2 and a newly-expanded set of injections. 

We structure this paper as follows: In Section~\ref{sec:injections} we describe the CHIME/FRB injection pipeline and the generation of a large population of synthetic bursts. In Section~\ref{sec:selection-correction-marginal} we present a procedure for fitting a fiducial model for the intrinsic FRB observable distributions to the marginalized distributions of observed quantities, using an importance-sampling procedure similarly applied by \citet{2021ApJS..257...59C}. In Section~\ref{sec:selection-function-modeling} we use the injections to fit a logistic regression model for the selection function. In Section~\ref{sec:refinement-scattering-distribution} we refine the estimate of the intrinsic scattering timescale distribution using a simulation-based inference procedure to better constrain the behavior of the distribution at large scattering times. In Section~\ref{sec:discussion} we compare the resulting debiased distributions to previous work and to results from other FRB surveys, and in Section~\ref{sec:conclusion} we present our conclusions. 

\section{CHIME/FRB Injections} \label{sec:injections}

This analysis requires a large catalog of synthetic FRB events injected into the CHIME/FRB real-time intensity data stream. These injections are processed by the same real-time detection, classification, and triggering pipeline used to identify astrophysical bursts, enabling an empirical characterization of the CHIME/FRB selection function. The injection framework used in this analysis was originally introduced by \citet{Merryfield}, who included an overview of the synthetic pulse generation pipeline and database networking structure that remain unchanged since the \citet{2021ApJS..257...59C} deployment. Here we summarize the architectural and population-level details most relevant to the present analysis.

\subsection{Injection Pipeline Architecture}

The injection system described in \citet{Merryfield} is designed to forward-model synthetic FRB signals through the CHIME/FRB real-time search pipeline, enabling direct measurement of survey completeness as a function of intrinsic burst properties. The CHIME/FRB search pipeline itself is implemented as a hierarchical, multi-stage software stack operating on beamformed intensity data generated by the CHIME F/X-engine (correlator and beamformer), as described in detail in the CHIME/FRB system overview \citep{CHIMEFRBSystemOverview}. In this architecture, signal processing proceeds through four principal stages, denoted L1 through L4, following an initial beamforming and upchannelization stage (L0). At L0, voltage data from the CHIME antennas are beamformed into 1024 static, FFT-formed intensity beams spanning the primary field of view, with 16,384 frequency channels at 0.983~ms cadence \citep{CHIMEFRBSystemOverview}. These intensity streams are distributed across a dedicated L1 compute cluster, with each L1 node processing data from eight static beams. The L1 pipeline conducts per-beam radio-frequency interference (RFI; see \citet{2023ApJS..265...62R}) excision and executes an efficient search for dispersed bursts, identifying candidate events in the time--DM plane using a highly optimized tree-dedispersion algorithm. Candidate detections from all beams are then consolidated at L2, where events coincident in time, dispersion measure, and sky position are grouped and refined using multi-beam information. L3 classifies grouped events as Galactic or extragalactic, cross-matches against a database of known FRB sources to identify candidate repeat events, and determines appropriate actions such as alerts or data callbacks; L4 then implements these actions and archives event metadata and associated data products for offline analysis \citep{CHIMEFRBSystemOverview}. Following successful triggering and data callback, candidate events are sent for manual inspection to determine whether they are astrophysical in origin. Inclusion in the final CHIME/FRB catalog requires independent confirmation by two human classifiers. 
\begin{deluxetable*}{lccccccc}[t]
\tablecaption{Summary statistics of the CHIME/FRB Catalog 2 sample after applying cuts (see Section \ref{sec:selection-correction-marginal}).
\label{tab:cat2-summary}}
\tablewidth{0pt}
\tablehead{
\colhead{Measured parameter} & \colhead{Median} & \colhead{Q25} & \colhead{Q75} & \colhead{Mean} & \colhead{SD} & \colhead{Min} & \colhead{Max}
}
\startdata
DM (pc cm$^{-3}$) & 558.52 & 378.14 & 832.21 & 652.88 & 390.68 & 122.524 & 3200.12 \\
Pulse width (ms) & 0.57 & 0.30 & 1.21 & 1.12 & 1.72 & 0.030 & 26.65 \\
Scattering timescale (ms) & 1.77 & 0.64 & 5.36 & 5.33 & 13.18 & 0.016 & 234.53 \\
\enddata
\tablecomments{Scattering timescales are referenced to 600 MHz after scaling the Catalog 2 measurements from 400 MHz assuming $\tau \propto \nu^{-4}$. Only a subset of bursts have measured scattering timescales from the \texttt{fitburst} dynamic spectrum fitting framework and may be treated as upper limits at the survey time resolution (0.983 ms) at 400 MHz \citep{Fitburst}.}
\end{deluxetable*}

Injections interface directly with the L1 stage, ensuring that synthetic bursts are embedded within the same RFI environment and downstream automated classification scheme as astrophysical events, while maintaining strict isolation between synthetic and real data products. The injection pipeline does not model vetting by humans, though any effect due to misclassification by humans is assumed to be minimal. The injections are coordinated through a centralized application programming interface (API) that interfaces directly with the L1 pipeline nodes. Each injection request specifies the burst properties, injection time, and target synthesized beam, and is assigned a unique identifier that propagates through the pipeline. Detection metadata produced at L1 and subsequent processing stages are matched to active injections using a database-backed bookkeeping system. This architecture enables a binary classification of each injected burst as detected or not detected and prevents spurious triggers from RFI or astrophysical events from being misidentified as injections, as detailed by \citet{Merryfield}.

\subsection{Sample Construction}
\label{sec:sample-construction}

A key requirement for the injected population is to span the parameter space of observed FRB properties characterized by fluence $F$, pulse width $w$, scattering timescale $\tau$, dispersion measure DM, and sky position (R.A., Decl.) broadly enough to characterize the survey response while avoiding an injection set overwhelmingly dominated by non-detections. Fluence and sky position are the principle drivers of detectability: naively sampling across these quantities would produce a detection fraction for injections that is prohibitively small. Accordingly, the sampling of sky positions must account for both the fiducial response of the primary and synthesized beams and the source position within an individual static beam. We therefore adopt a hierarchical sampling strategy to construct an injection sample that balances exploration of parameter space with efficiency. This procedure follows in three stages:
\begin{enumerate}
\item \textbf{Initial sampling of catalog observables.}  We draw DM, intrinsic width, and scattering timescale from lognormal distributions informed by the observed catalog, supplemented by a uniform exploration fraction extending beyond the catalog distributions. We provide a table of summary statistics for each distribution in Table \ref{tab:cat2-summary}. For each of these parameters, $70\%$ of injections are drawn from lognormal distributions fit to the Catalog 2 distributions. The remaining $30\%$ are drawn from uniform distributions extending well beyond these observed parameter distributions. This increases sampling density in the regime where the detection probability is expected to transition from near unity to near zero.
\item \textbf{Fluence and sky-position sampling.}  Fluences are drawn from a power-law distribution, $p(F) \propto F^{-1.5}$, corresponding to the expectation for a Euclidean, non-evolving source population tentatively evidenced by \citet{2021ApJS..257...59C}. Relative to the injection campaign accompanying Catalog~1, we adopt slightly enhanced sampling at the low-fluence end to increase the density of injections near the detection threshold. This improves the resolution with which the transition in detection probability can be constrained. The lower fluence bound is set at 0.1 Jy ms, near the approximate noise floor of CHIME/FRB detections and well below the estimated survey $95\%$ completeness threshold of $\sim\!5$ Jy ms \citep{Catalog2}, so that a non-negligible fraction of injections populate the regime of near-zero detection probability; the upper bound extends to $\gtrsim10^4$ Jy ms, well beyond the brightest cataloged events. 

CHIME/FRB does not have a single, well-defined fluence threshold; rather, the effective detection threshold varies across the sky as a function of beam sensitivity. Accordingly, sky positions\footnote{In contrast to catalog events which each have measured header localizations, each injection is assigned a simple coordinate within the CHIME beam model for which a directional sensitivity can be calculated.} are drawn uniformly on the celestial sphere and initially filtered to exclude locations below the true horizon at CHIME's latitude. The accepted positions are then mapped through the CHIME primary beam model to compute the gain and effective sensitivity at the time of injection, which is highly frequency dependent across the CHIME band \citep[see][]{2023AJ....166..138A}.  Positions with a primary beam response\footnote{The beam model presented in, e.g, \citet{2022ApJ...932..100A} defines the primary beam sensitivity as normalized at the transit of calibrator source Cygnus~A for all frequencies, and the FFT-formed synthesized beams as normalized relative to the on-axis sensitivity.} below $10^{-3}~{\rm Jy/Jy}$ and synthesized beam response below $10^{-2}~{\rm Jy/Jy}$ are rejected to avoid scheduling injections in regions of negligible sensitivity relative to the main lobe.

\item \textbf{Pre-selection using beam sensitivity and estimated S/N.} For each candidate injection, we estimate a lower bound on the expected S/N by using a randomly drawn sky position from the previous step and computing its attenuation of the burst dynamic spectrum using a model of the CHIME/FRB beam response \citep[e.g.][]{2022ApJ...932..100A, 2023AJ....166..138A}. Injections predicted to have an estimated lower-limit ${\rm S/N} < 12$ under all beam realizations are rejected prior to scheduling. This step reduces the computational overhead of injecting many otherwise-undetectable bursts while preserving sampling in the vicinity of the survey threshold.
\end{enumerate}

\begin{figure}[t]
\centering
\includegraphics[width=0.95\textwidth]{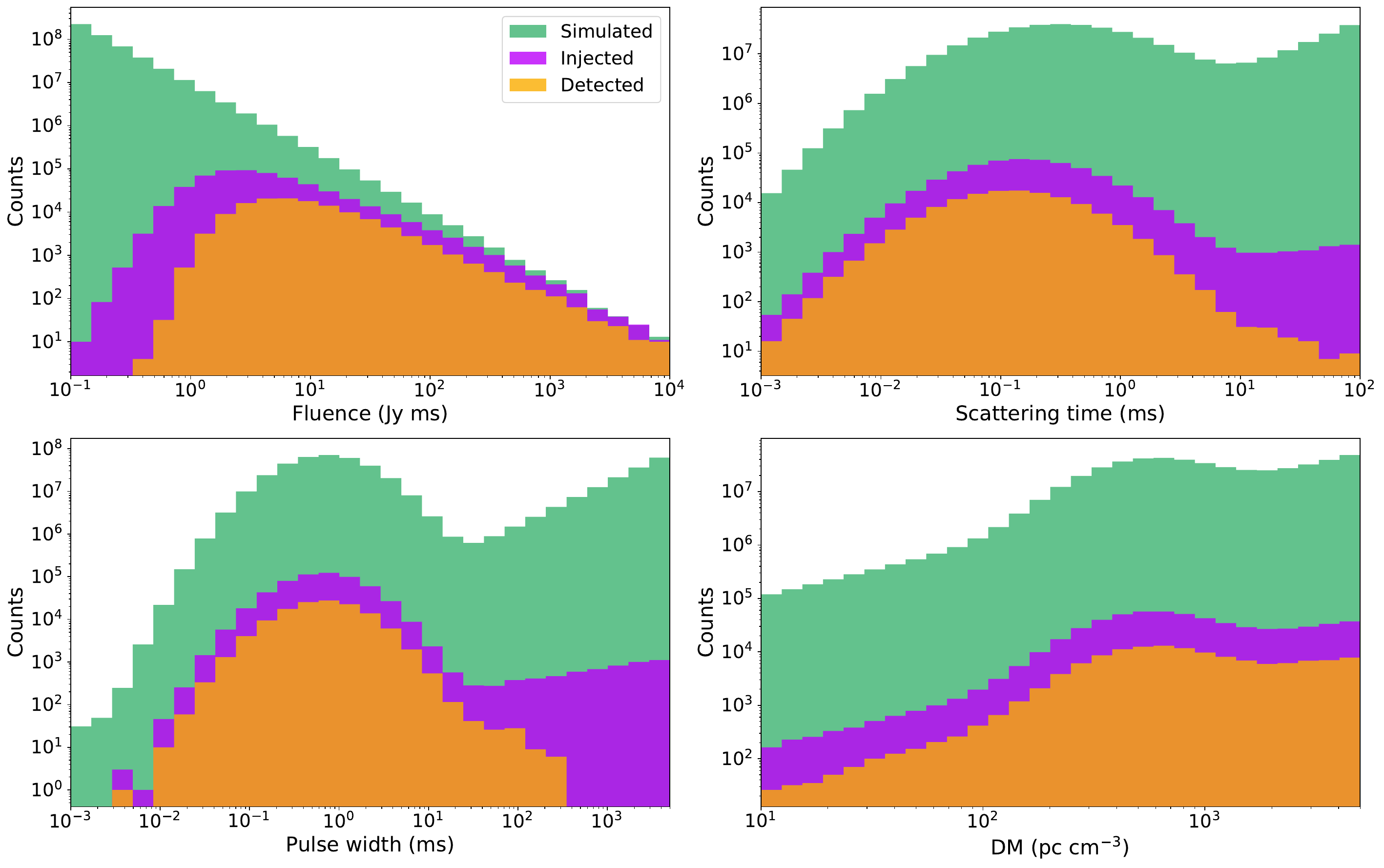}
\caption{Marginalized distributions of the simulated (green), injected (purple), and detected (orange) populations for the injection campaign. Each panel shows one-dimensional projections in fluence, scattering timescale (referenced to 1 GHz), intrinsic pulse width, and dispersion measure (DM). The simulated population represents the underlying draws from distributions widely sampling the FRB parameter space, while the injected subset reflects the samples pre-selected for injection (see Section \ref{sec:sample-construction}), and the detected sample corresponds to injected events with corresponding detection.}
\label{fig:inj-marginals}
\end{figure}

We show the resulting marginalized distributions of simulated, injected, and detected events in Figure~\ref{fig:inj-marginals}. The detected subset is biased toward higher fluence and smaller effective widths, consistent with the single-pulse detection metric ${\rm S/N} \propto {F}~/\sqrt{w_{\rm obs}}$, where $w_{\rm obs}$ is the quadrature sum of intrinsic width, intra-channel dispersion smearing, and scatter broadening \citep{2003ApJ...596.1142C}. As such, broader bursts are more difficult to detect at fixed fluence when the width is greater than the 0.983 ms time resolution. The injected population is otherwise consistent with that employed for Catalog~1 \citep{Merryfield}, with several extensions: (i) an increased uniform exploration fraction for DM, intrinsic width, and scattering timescale (from $10\%$ to $30\%$); (ii) an expanded upper bound on intrinsic width extending to 5 seconds to support concurrent injections for the CHIME/Slow backend\footnote{CHIME/Slow is a commensal search pipeline designed to detect long-duration radio transients with widths from 16 milliseconds up to 5 seconds, and has been detecting events in this regime from October 2025 onwards.} \citep{CHIMESlow}; and (iii) enhanced low-fluence sampling to improve characterization of the survey threshold. 

Scattering timescales are parameterized at a reference frequency of 600~MHz and scaled across the band assuming $\tau_{\rm sc} \propto \nu^{-4}$, corresponding to a Kolmogorov-like scattering index; this convention is adopted for all scattering timescales in this paper unless otherwise stated. At present, injections are limited to single-component bursts, both due to constraints in the dynamic spectrum simulator and as a simplification consistent with the predominance of apparently single-component broadband events detected by CHIME/FRB \citep{ZiggyRepeaterMorphology}. These injections therefore assume a Gaussian intrinsic pulse profile convolved with a one-sided exponential scattering tail. This does not capture the full morphological complexity of real FRBs, notably multi-component structure, asymmetric temporal profiles, and non-exponential scattering tails. However, as CHIME/FRB uses boxcar matched filtering on dedispersed intensity data, this parameterization is broadly applicable to the characterization of most observed FRBs by CHIME/FRB's search pipeline. While we do not model selection effects due to spectral morphology, we sample the parameters of spectral index and spectral running, that describe the slope and curvature of the burst spectrum across the observing band, respectively, from a 2D kernel density estimate in constructing the population, following \citet{2021ApJS..257...59C}. This allows for marginalization across these parameters in the subsequent analysis.

\subsection{September 2025 Injection Campaign}

The injection dataset used in this work was collected during a dedicated campaign conducted from 2025 September 8--12. The campaign was designed to maximize injection throughput while maintaining stable pipeline behavior, which would alternatively cause a high rate of RFI masking of injections if the injection rate is too high. Additionally, during this period all downstream pipelines not required for injection tracking were temporarily disabled out of caution for these systems operating alongside the CHIME/FRB backend. This included intensity and baseband callbacks, VOEvent generation \citep[see][]{2025AJ....169...39A}, and most L4 processing and database services. These components were restored to normal operation immediately following the conclusion of the run. In total, the campaign produced 587,367 injections, of which 130,763 have corresponding detections. The marginalized property distributions of the simulated, injected, and detected events are shown in Figure~\ref{fig:inj-marginals}.

\section{Correcting for marginalized selection effects}
\label{sec:selection-correction-marginal}

We present here the procedure for fitting a fiducial model of the debiased FRB population, closely following the resampling-based formalism introduced in Section~6 of \citet{2021ApJS..257...59C}. While the full methodology is developed in that work, we summarize the relevant components here as they pertain to the present analysis. We restrict analysis to fluence, DM, intrinsic width $w$, and scattering timescale $\tau$, and defer a detailed treatment of burst morphologies to a future dedicated injection campaign. A central motivation for this work is determining the intrinsic distribution of scattering timescales, which remained only weakly constrained in Catalog~1 owing to limited catalog statistics at moderate and large scattering times ($\tau \gtrsim 10\,\mathrm{ms}$ at 600~MHz) and comparatively sparse injection coverage in this regime. Constraining this distribution is important as scattering traces turbulent plasma along the line of sight of the burst, and prior population modeling has shown that the observed scattering distribution by CHIME/FRB is more consistent with FRBs originating in dense local environments compared to the average host-galaxy ISM alone \citep[e.g.,][]{Pragya}. 

We implement several cuts to the Catalog 2 and injection samples to ensure a consistent and well-defined comparison between the observed population and the injections. For the catalog, we remove events affected by known exposure issues (including pre-commissioning data) as well as bursts detected in sidelobes ($>5$ degrees from the meridian). We further impose cuts in S/N ($\mathrm{S/N} \geq 12$) and DM ($\mathrm{DM} \geq 100\ \mathrm{pc\ cm^{-3}}$). The S/N cut is intended to mitigate incompleteness associated with human verification near the triggering threshold ($\mathrm{S/N} \geq 8$), while the DM cut prevents contamination from Galactic sources. We also require the fitburst width to exceed $3\times10^{-2}$ ms, matching the lower bound of the injected width distribution (see Figure \ref{fig:inj-marginals}). Additionally, we remove known repeat bursts by retaining only the first detection from each source  \citep[as identified by][]{RN4}. Ignoring measurements from subsequent repeat events prevents highly active repeating sources from dominating the sample and ensures the same selection process for repeaters as apparent one-offs.

We also remove bursts observed along sightlines where the Galactic contribution to scattering is expected to be significant. Using the NE2025 model \citep{2026ApJ..1002....3O}, we exclude bursts for which the predicted Milky Way scattering exceeds the CHIME/FRB search time resolution of 0.983 ms at 400 MHz. This cut removes events along sightlines for which the observed scattering could plausibly be dominated by the Galactic plane, rather than by propagation local to the source, its host galaxy, the circumgalactic medium, or the intergalactic medium. While the Catalog 2 scattering timescale distribution does not strongly depend on Galactic latitude, we nevertheless remove sightlines where the Milky Way contribution to scattering is expected to be significant. This cut removes fewer than $5\%$ of the remaining sample, so Galactic scattering is unlikely to be the dominant driver of the observed scattering distribution. 

\begin{figure}[t]
    \centering
    \includegraphics[width=\linewidth]{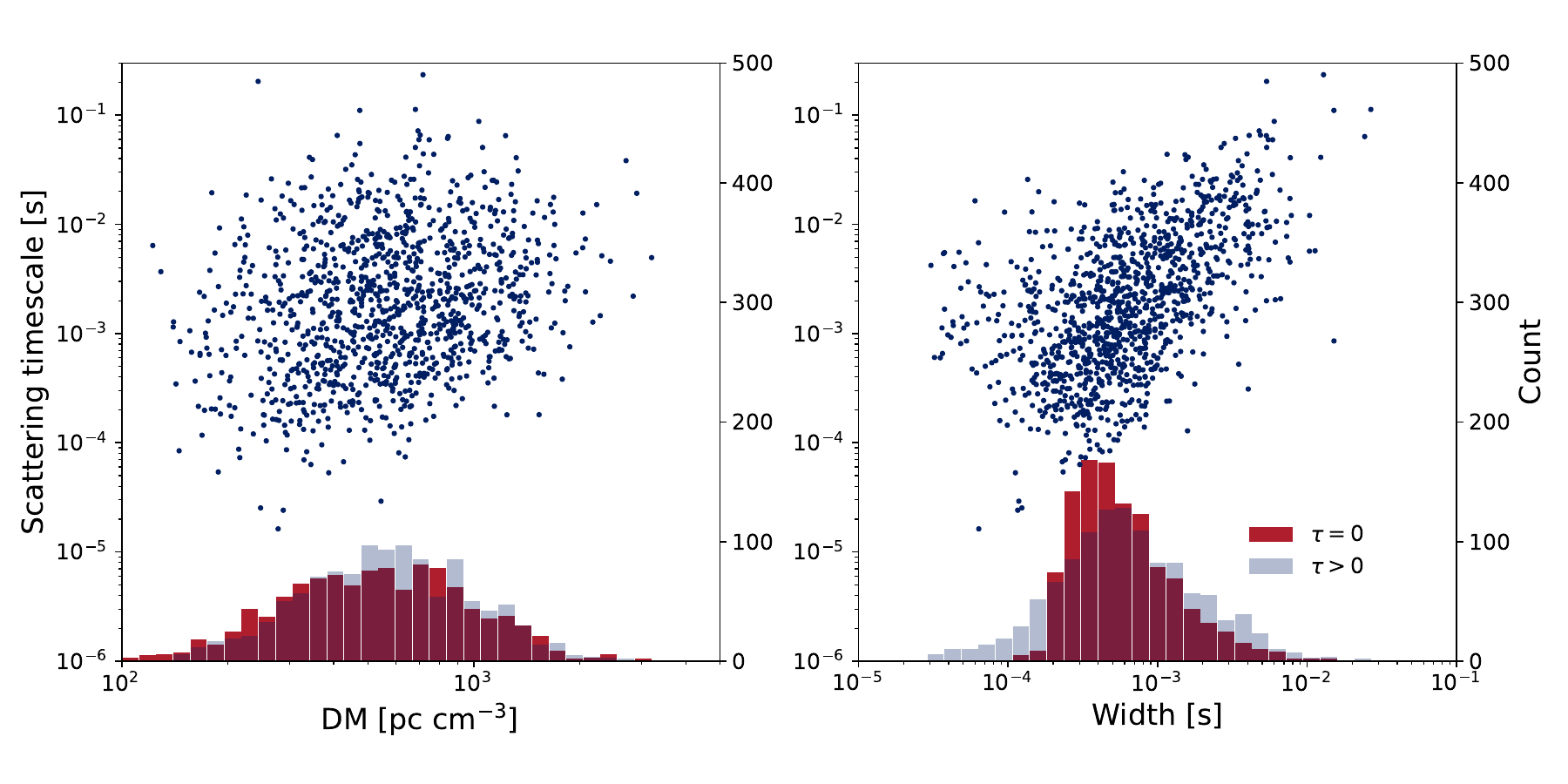}
    \caption{Scattering timescale as a function of DM and width for the Catalog~2 sample with cuts applied (see Section \ref{sec:selection-correction-marginal}). Blue points show individual events. Background histograms compare the distributions of events that do not prefer a scatter-broadened profile ($\tau=0$) and events with measured scattering timescales ($\tau>0$). }
    \label{fig:cat2-scatULs-withfluence}
\end{figure}

We also exclude all bursts without measured scattering timescales from the analysis. In Catalog~2, events with reported $\tau=0$ do not represent measurements of zero scattering, but cases in which \texttt{fitburst} prefers an unscattered Gaussian profile over a scatter-broadened profile \citep{Fitburst}. This may be due to an intrinsically very low scattering time, or that the intrinsic pulse width dominates over an otherwise measureable scattering value. These events therefore provide upper limits on $\tau$, with the limit canonically set at twice the observed pulse width; previous population analyses have applied these events by assigning them an effective scattering timescale set at half the upper limit, i.e., $\tau=w$ \citep[see][]{2021ApJS..257...59C, ShinEtAl}. However, we do not perform the same procedure, as this may artificially concentrate probability mass within an otherwise measureable part of parameter space, particularly as the true scattering timescales of these events may be much lower. Figure~\ref{fig:cat2-scatULs-withfluence} shows that the $\tau=0$ and $\tau>0$ samples occupy broadly similar ranges of DM and width. Thus, we excise the sample with unmeasured scattering times as it does not strongly bias the intrinsic distributions of these properties. 

For the scattering timescale distribution itself, however, we caution that this analysis is more sensitive to the high-scattering end of the distribution, where scattering can be well-measured, rather than the low-scattering end. Forthcoming CHIME/FRB data products with baseband data are better suited to address this limitation for Catalog 2, and such work is currently underway (Amouyal et al. in prep.). These data preserve raw complex voltages from most intensity triggers with $\mathrm{S/N} \geq 12$ and give an improved time resolution of 2.56 $\mu$s, enabling improved morphological characterization of low-scattering events \citep[e.g.][]{Basecat1, 2025ApJ...979..160S}. As such, we defer a detailed study of the low-scattering sample to future work. These cuts to the catalog sample give a total of 1079 bursts with which we apply the following analysis. 

\subsection{Fiducial Model}
The intrinsic FRB population is described by a parametric probability density $p(F, \mathrm{DM}, w, \tau)$. Here, $w$ represents an intrinsic Gaussian pulse width. As an initial model, we adopt a separable fiducial model for the underlying population,
\begin{equation}
p(F, \mathrm{DM}, w, \tau)
= p(F \mid \alpha)\,
  p(\mathrm{DM} \mid \mu_{\mathrm{DM}},\, \sigma_{\mathrm{DM}})\,
  p(w \mid \mu_w,\, \sigma_w,)\,
  p(\tau \mid \mu_\tau,\, \sigma_\tau),
\end{equation}
where $\alpha$ is the fluence power law index and $\{\mu,\, \sigma \}$ represent the shape and scale of a lognormal distribution. The intrinsic fluence distribution is therefore modeled: 
\begin{equation}
p(F \mid \alpha)
= \frac{\alpha - 1}{F_{\min}}
\left( \frac{F}{F_{\min}} \right)^{-\alpha},
\qquad F \ge F_{\min},
\end{equation}
with $F_{\min}$ fixed at $0.1\,\mathrm{Jy\,ms}$, well below the survey completeness limit (see Figure~\ref{fig:inj-marginals}). This gives a fiducial approximation of the fluence distribution over the range probed by CHIME/FRB, and in reality will not extend to indefinitely high burst energies. A turnover in the energy distribution is expected above $E_{\max}\sim10^{42}\,\mathrm{erg}$ \citep{2026MNRAS.545f1937O, 2026arXiv260219335S}, and observations of hyperactive repeating sources show that burst energy distributions can deviate from a simple power law \citep[e.g.,][]{Kirsten2024,OuldBoukattine2026a,OuldBoukattine2026b}.

The intrinsic width and scattering timescale distributions are each modeled as a log-normal distribution in $\ln w$ and $\ln \tau$, respectively. For an observable $x \in \{w, \tau, \mathrm{DM}\}$, the log-normal probability density is
\begin{equation}
p(x \mid \mu_x, \sigma_x)
=
\frac{1}{x\,\sigma_x\sqrt{2\pi}}
\exp\!\left[
-\frac{(\ln x - \mu_x)^2}{2\sigma_x^2}
\right],
\qquad x > 0 .
\end{equation}
The DM distribution is treated phenomenologically using this log-normal parameterization matched to the injected population, without explicitly decomposing Galactic, host, and intergalactic contributions.
\begin{figure*}[t]
    \centering
    \includegraphics[width=\linewidth]{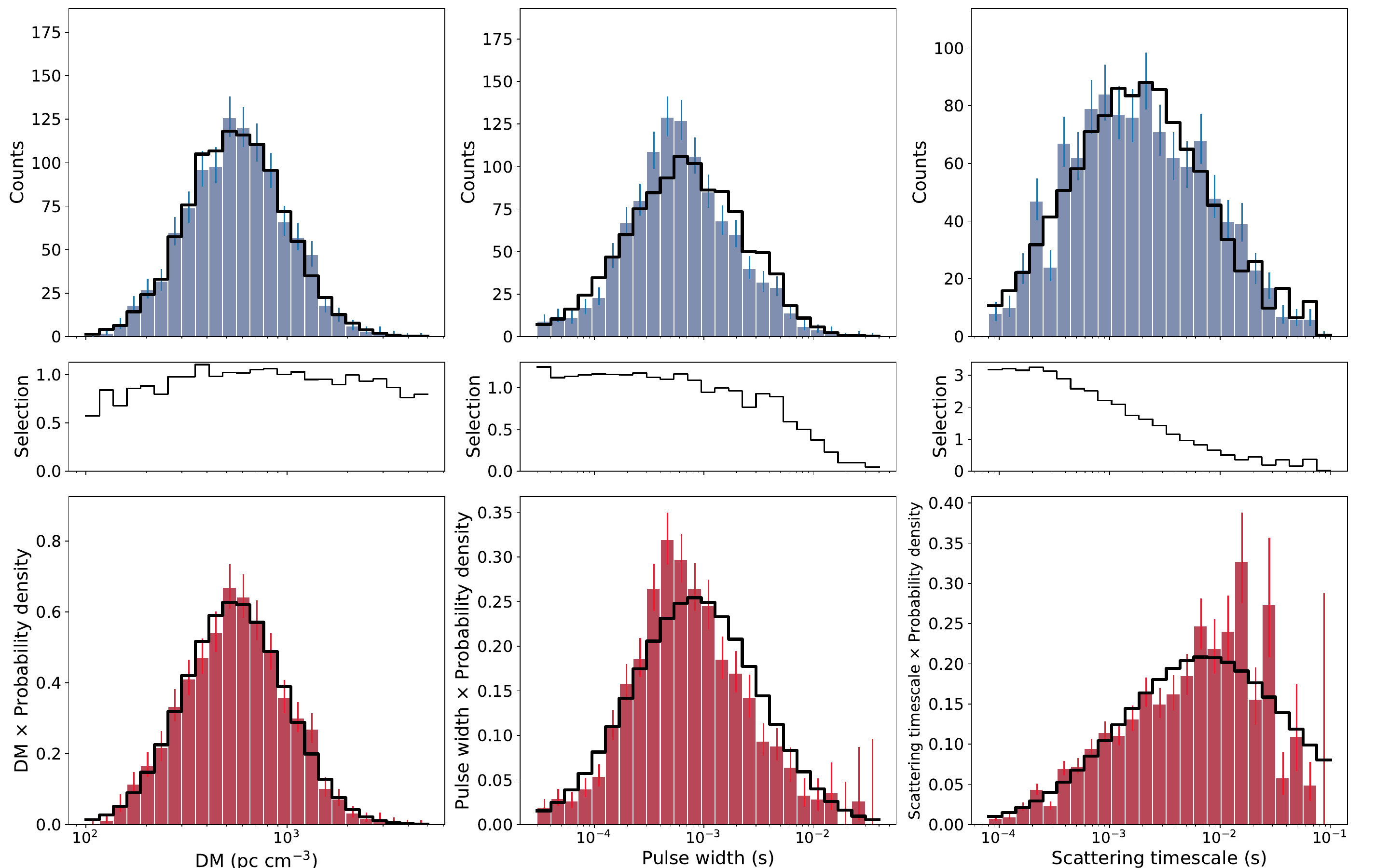}
    \caption{Binned selection plots for DM, pulse width and scattering time. (Top) Catalog 2 DM, pulse width, and scattering timescale distributions (bars), with a fiducial model modulated by a selection effect (solid lines). (Middle) Relative marginalized selection curve inferred during fiducial model fitting. (Bottom) Selection-corrected observed distribution (bars) with fiducial model (solid lines).}
    \label{fig:width_dm_binned}
\end{figure*}
The full hyperparameter vector may therefore be written as
\begin{equation}
\boldsymbol{\phi}
= \{\alpha,\,
\mu_{\mathrm{DM}},\, \sigma_{\mathrm{DM}},\,
\mu_w,\, \sigma_w,\,
\mu_\tau,\, \sigma_\tau \}.
\end{equation}
This parameterization provides a compact and interpretable description of the intrinsic FRB population while remaining sufficiently flexible to capture the broad features of the observed catalog when correcting for selection effects.

Following Catalog~1, population inference is not performed directly in fluence space. CHIME/FRB intensity fluences are best interpreted as lower limits due to uncertainty in localization within the synthesized beam \citep{2023AJ....166..138A}. The analysis is therefore carried out in the space of DM and observed signal-to-noise ratio (S/N), which are highly correlated with fluence \citep[see][]{2021ApJS..257...59C} and robustly measured for all catalog events. The injection dataset is used to calibrate the instrument response by forward-modeling the detection probability as a function of intrinsic burst properties, implicitly encoding the mapping between $(F, \mathrm{DM}, w, \tau)$ and the observed S/N. The predicted observable distribution in $(\mathrm{DM}, \mathrm{S/N})$ space is therefore obtained by integrating over fluence,
\begin{equation}
P_{\mathrm{obs}}(\mathrm{DM}, \mathrm{S/N})
= \int \mathrm{d}F\,
P(F, \mathrm{DM})\,
P(\mathrm{S/N} \mid F, \mathrm{DM}),
\end{equation}
allowing the intrinsic fluence distribution to be constrained indirectly through its imprint on the observed S/N distribution, while avoiding uncertain beam corrections to individual fluence measurements that the injection system does not robustly inverse-model.

The population parameters are constrained using an iterative maximum-likelihood fitting procedure identical to that used by \citet{2021ApJS..257...59C}. At each iteration, the predicted catalog distributions are constructed by reweighting the detected injections according to the ratio between the current population model and the injected reference distribution. The resulting weighted injection sample provides a forward-modeled prediction for the observed catalog distributions in DM, intrinsic width, scattering timescale, and S/N. Model parameters are then updated by fitting these predicted distributions to the observed catalog via the binned Poisson likelihood,
\begin{equation}
\ln \mathcal{L}(\boldsymbol{\phi}) = \sum_i \left[ n_i \ln \mu_i(\boldsymbol{\phi}) - \mu_i(\boldsymbol{\phi}) \right],
\end{equation}
where $n_i$ is the binned catalog count in bin $i$ and $\mu_i(\boldsymbol{\phi})$ is the reweighted injection prediction. The fluence power-law index is constrained through the DM--S/N distribution. In Figure~\ref{fig:width_dm_binned} we present the fit results of the fiducial population model to the observed catalog distributions. We also provide the parameter estimates for $\phi$ in Table~\ref{tab:cat1_cat2_params}, including bootstrapped $95\%$ confidence intervals on each parameter. Note that, for DM and intrinsic width, the majority of catalog events lie in regions of relatively high selection probability, and the corresponding selection-corrected distributions are therefore well constrained. In contrast, the scattering timescale distribution extends substantially into a regime of low selection probability. When corrected using inverse-probability weighting, this behavior implies a comparatively large contribution from intrinsically highly scattered events that is increasingly dominated by statistical noise. In Catalog~1, this regime was similarly dominated by Poisson counting uncertainties owing to the small number of detected bursts at moderate and large scattering times. This effect is amplified due to choice of binning. Accordingly, \citet{2021ApJS..257...59C} quoted an upper limit of 10 ms for the well-characterized domain of the scattering-timescale distribution. We therefore proceed with a kernel density estimation (KDE) procedure that allows us to analogously define the upper limit of the scattering timescale distribution well-characterized by Catalog 2. 
\begin{deluxetable*}{lccccccc}
\tablecaption{Best-fit fiducial model parameters with 1$\sigma$ confidence intervals from analogous cuts and fitting procedure with the CHIME/FRB Catalog~1 sample \citep{2021ApJS..257...59C} and Catalog~2 \citep{Catalog2}.
\label{tab:cat1_cat2_params}}
\tablewidth{0pt}
\setlength{\tabcolsep}{3pt}
\tablehead{
\colhead{} &
\colhead{$\alpha$} &
\colhead{$\mu_{\rm DM}$ [pc cm$^{-3}$]} &
\colhead{$\sigma_{\rm DM}$} &
\colhead{$\mu_{w}$ [s]} &
\colhead{$\sigma_{w}$} &
\colhead{$\mu_{\tau}$ [s]} &
\colhead{$\sigma_{\tau}$}
}
\startdata
Catalog~1 &
$-1.35 \pm 0.12$ &
$494 \pm 24$ &
$0.67 \pm 0.04$ &
$(9.80 \pm 0.91)\times10^{-4}$ &
$0.97 \pm 0.08$ &
$(2.15 \pm 1.10)\times10^{-3}$ &
$1.71 \pm 0.23$ \\[4pt]
Catalog~2 & 
$-1.07 \pm 0.02$ & 
$548 \pm 4$ & 
$0.58 \pm 0.01$ & 
$(8.39 \pm 0.35)\times10^{-4}$ & 
$1.34 \pm 0.03$ & 
$(7.49 \pm 0.60)\times10^{-3}$ & 
$1.79 \pm 0.04$ \\
\enddata
\tablecomments{
Uncertainties are bootstrap-derived. Shape ($\mu$) and scale ($\sigma$) parameters describe the dispersion measure (DM), width ($w$), and scattering timescale ($\tau$) distributions. Catalog~1 and Catalog~2 use different treatments of scattering upper limits: Catalog~1 assigns unmeasured scattering timescales to the burst width, while Catalog~2 excludes bursts without measurable scattering from the analysis (see text).}
\end{deluxetable*}
\subsection{Nonparametric Density Estimation}

To mitigate the sensitivity of the selection-corrected distributions to binning choices, we additionally estimate the intrinsic distributions using a KDE for nonparametric density. This is particularly useful for the scattering timescale distribution, which displays a modest number of detections in regions with large selection bias (see Figure~\ref{fig:width_dm_binned}). While we find that this approach yields smooth estimates for DM and intrinsic width, the Gaussian-kernel estimate for scattering exhibits an apparent drop-off at large $\tau$. Given the strongly selection-limited nature of this regime and the small number of contributing catalog events, this behavior may not necessarily reflect a physical drop-off in the intrinsic scattering distribution, and instead arises from the asymptotic behavior of the reweighted kernel in regions of sparse data.

\begin{figure*}
    \centering
    \includegraphics[width=0.95\linewidth]{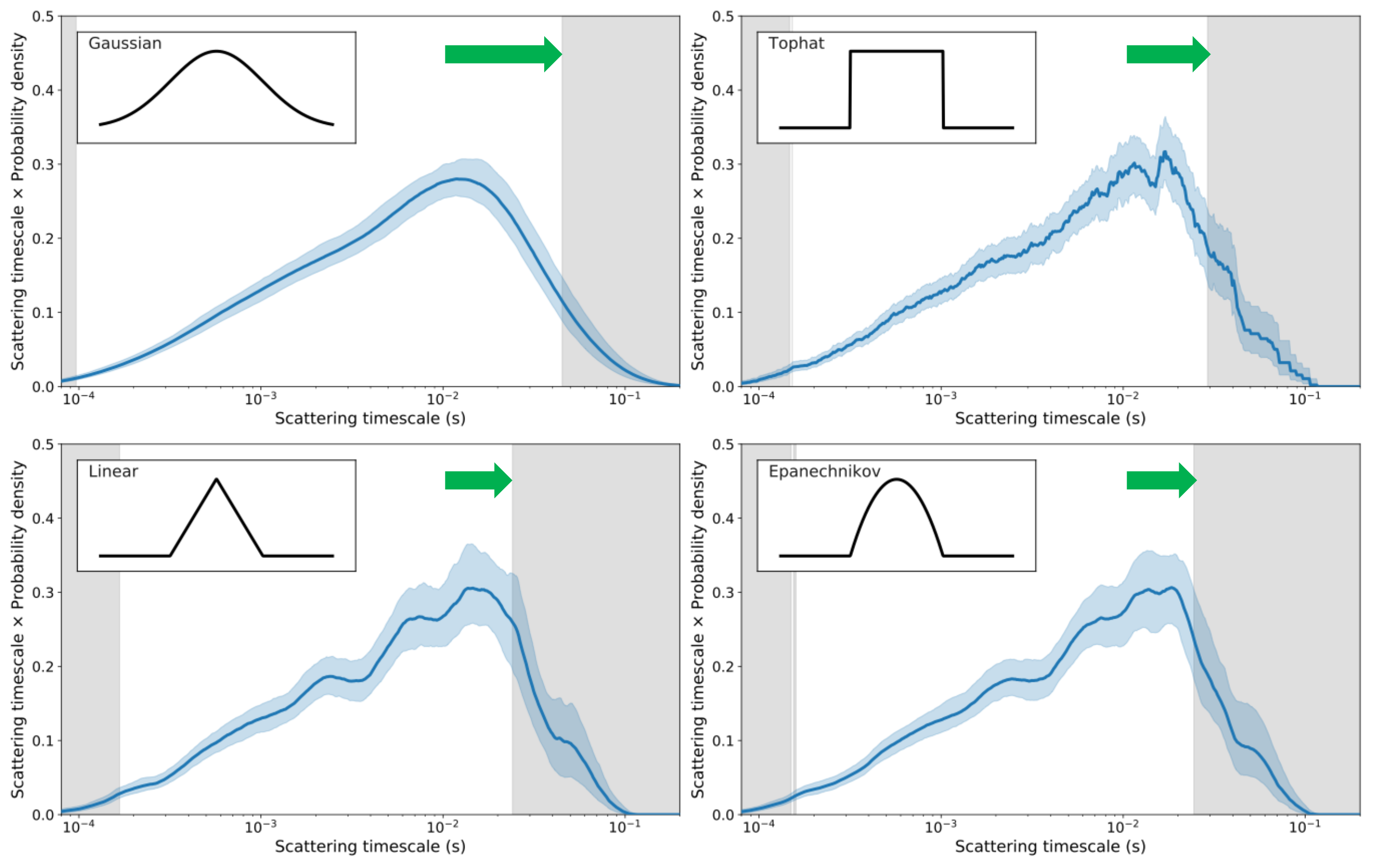}
    \caption{Selection-corrected scattering timescale distributions at 600~MHz inferred via KDE using four kernel choices (Gaussian, tophat, linear, and Epanechnikov; insets). Solid curves denote the bootstrap median KDE, with shaded regions indicating the 95\% confidence interval on the inferred distribution. The gray region marks the regime where the local dominance metric exceeds the threshold $d_{\mathrm{Cat1}}(10~\mathrm{ms}) = 0.122$ (see text surrounding Equation~\ref{eq:ldm}) calibrated from the Catalog~1 analysis, corresponding to a breakdown of reliable inference due to domination by a small number of heavily weighted events. This threshold occurs at $\tau \sim 30$ ms, where the local effective sample size is $n_{\rm eff} \approx 50$. Beyond this scale, the KDE is increasingly noise-dominated and should not be interpreted as physically constraining. The onset of this noise-dominated regime is indicated by the left edge of the gray shaded region. The increased range relative to Catalog~1 for each kernel choice is indicated by the green arrows.}
    \label{fig:scat-time-kernels}
\end{figure*}

As we expect CHIME/FRB observations may only probe the low-$\tau$ tail of a broader intrinsic scattering distribution, consistent with results from other surveys that do not show a clear high-$\tau$ cutoff,\citep[e.g.,][]{2026PASA...43...38J}, we therefore explore alternative kernel choices for the scattering KDE. Specifically, we construct selection-corrected KDEs using boxcar, triangular, and Epanechnikov kernels \citep[see][]{scikit-learn}, all of which fall identically to zero outside their respective bandwidths. The KDE distributions are shown in Figure~\ref{fig:scat-time-kernels} with their respective kernels. We apply Scott's rule for bandwidth estimation \citep{10.1093/biomet/66.3.605}, computing the kernel in $\log_{10}\tau$ space. These alternative estimates more transparently illustrate the uncertainty associated with extrapolating beyond the range strongly constrained by the catalog. Uncertainties on the KDEs are quantified using a Bayesian bootstrap on the inverse-probability weights from which we derive a 95\% confidence interval on the inferred distributions \citep{rubin1981bayesian}. Specifically, we treat the selection-corrected sample as a weighted empirical distribution and generate bootstrap realizations by drawing from a Dirichlet distribution to perturb the normalized weights. This technique is convenient as we can resample the probability mass assigned to each burst while holding the observed $\tau$ values fixed and without rerunning the computationally expensive fiducial model fitting procedure. The resulting ensemble of reweighted samples is then propagated through the KDE estimator to obtain bootstrapped confidence intervals that reflect finite-sample uncertainty in the selection-corrected distribution, conditioned on the adopted selection function model.

To identify the upper limit of the selection-corrected scattering time distribution that remains meaningfully constrained by the data, we introduce two diagnostics: the local effective sample size, \(n_{\mathrm{eff}}\), and a local dominance metric, \(d(\tau)\). While the selection-corrected KDE exhibits a decline in the inferred intrinsic scattering distribution above \(\sim 50\,\mathrm{ms}\), this feature may be caused by a combined consequence of severe selection bias and the influence of the smoothing kernel, even when alternative kernel choices (Figure~\ref{fig:scat-time-kernels}) are considered. Based on the design effect defined by \citet{kish1965survey}, we hence define the effective sample size
\begin{equation}
\label{eqn:tau_neff}
    n_{\mathrm{eff}}(\tau) = \frac{\left(\sum_i w_i K_h(\tau - \tau_i)\right)^2}{\sum_i \left(w_i K_h(\tau - \tau_i)\right)^2},
\end{equation}
and local dominance metric
\begin{equation}
    d(\tau) = \frac{\max_i \left\{ w_i K_h(\tau - \tau_i) \right\}}{\sum_j w_j K_h(\tau - \tau_j)}.
    \label{eq:ldm}
\end{equation}
Here, \(K_h\) is the kernel function evaluated with bandwidth \(h\), and \(w_i = 1/S(\tau_i)\) represents the selection-corrected weight for each detection. The term \(w_i K_h(\tau - \tau_i)\) quantifies how much detection \(i\) contributes to the density estimate at \(\tau\), based on both its proximity to \(\tau\) and its inverse probability weight. The effective sample size \(n_{\mathrm{eff}}(\tau)\) gives the equivalent number of equally contributing detections required to produce the same level of statistical support. Meanwhile, the dominance metric \(d(\tau)\) captures the fractional influence of the most dominant detection. By placing a threshold on \(d(\tau)\), we define a limit on how much any single event may dominate the KDE. These metrics allow us to define a region beyond which the KDE is shaped primarily by kernel smoothing, and is thus uninformative.

To ensure a consistent definition of the regime in which the scattering timescale KDE is meaningfully constrained by the data and consistent with previous population inference, we calibrate these diagnostics using the results of \citet{2021ApJS..257...59C} and then apply the same thresholds to Catalog~2. For the Catalog~1 inference, we evaluate the local dominance metric $d(\tau)$ at $\tau = 10\,\mathrm{ms}$, which corresponds to the informative upper limit placed on the scattering distribution in the previous analysis. There we measure a dominance value of $d_{\mathrm{Cat1}}(10\,\mathrm{ms}) = 0.122$, which we adopt as a conservative threshold defining the maximum contribution of any single selection-corrected catalog event to the KDE. For Catalog~2 we compute the same diagnostics using an identical set of kernel distributions and bandwidth optimization procedures, identifying the largest scattering timescale $\tau_{\rm max}$ for which $d(\tau) \le d_{\mathrm{Cat1}}(10\,\mathrm{ms})$ and the local effective sample size remains non-zero. Applying this criterion, we find the selection-corrected scattering timescale distribution remains similarly informative up to $\tau \lesssim 30~\textrm{ms}$ with a corresponding local effective sample size of $n_{\rm eff}\approx 50$ at the threshold. Beyond this scale, the KDE is increasingly dominated by statistical noise due to a small number of weighted detections, and is not necessarily physically constraining.

\section{Explicit Selection Function Modeling} 
\label{sec:selection-function-modeling}

Using the injection data, we model the CHIME/FRB selection function as the probability that a burst is detected above a specified signal-to-noise threshold, analogous to placing a binary probability on the observation function defined by \citet{2021ApJS..257...59C}. We define
\begin{equation}
\label{eqn:SelectionFunctionDefinition}
    S(\mathbf{x}) =
    P({\rm S/N} > {\rm [S/N]}_{\rm thresh} \mid \mathbf{x}),
\end{equation}
where $\mathbf{x}=(F,\mathrm{DM},\tau,w)$ denotes the vector of observed burst parameters. While \citet{2021ApJS..257...59C} assumed that this high-dimensional detection probability factorizes into independent contributions from each variable, we instead model the selection function directly in the full four-dimensional parameter space using the injection set as labeled training data. This allows the detection probability to be inferred empirically without assuming separability.

We adopt a logistic regression framework to model the detection probability. For each burst, let $\boldsymbol{\phi}(\mathbf{x})$ denote the feature vector obtained from $\mathbf{x}$ after applying a chosen polynomial expansion, including interaction terms. Equivalently, for the full injection sample, the design matrix $\mathbf{X}$ is constructed by stacking the feature vectors $\boldsymbol{\phi}(\mathbf{x}_i)^T$ for all injected bursts. The selection function evaluated at a single point $\mathbf{x}$ is then modeled as
\begin{equation}
    S(\mathbf{x}\mid\boldsymbol{\beta})
    =
    \frac{1}
    {1+\exp[-\boldsymbol{\phi}(\mathbf{x})^{T}\boldsymbol{\beta}]},
\end{equation}
where $\boldsymbol{\beta}$ is the corresponding vector of regression coefficients. Equivalently, the log-odds of detection are given by
\begin{equation}
    \ln \left[
    \frac{S(\mathbf{x}\mid\boldsymbol{\beta})}
    {1-S(\mathbf{x}\mid\boldsymbol{\beta})}
    \right]
    =
    \boldsymbol{\phi}(\mathbf{x})^{T}\boldsymbol{\beta},
\end{equation}
which defines the logistic link between the probability scale and the linear predictor. For example, for a second-order polynomial expansion,
\begin{equation}
    \boldsymbol{\phi}(\mathbf{x})^{T}\boldsymbol{\beta}
    =
    \beta_0
    + \sum_{i=1}^{N} \beta_i x_i
    + \sum_{i=1}^{N}\sum_{j\ge i}^{N}\beta_{ij}x_i x_j,
\end{equation}
with higher-order terms included in turn. We apply various methods for model selection to determine the optimal order of the polynomial expansion (see Section~\ref{sec:model_selection}). By applying the delta method \citep[see, e.g.,][]{AllOfStatistics}, which uses a first-order Taylor expansion to propagate uncertainty in the fitted parameters under asymptotic normality, we approximate the variance of the selection function evaluated at $\mathbf{x}$, $\mathrm{Var}[S(\mathbf{x};\hat{\boldsymbol{\beta}})]$, from the covariance of the estimated coefficients, $\mathrm{Cov}(\hat{\boldsymbol{\beta}})$, thereby obtaining the variance on the fit:
\begin{eqnarray}
\mathrm{Var}\!\left[S(\mathbf{x};\hat{\boldsymbol{\beta}})\right]
&\approx&
\left[\nabla_{\boldsymbol{\beta}} S(\mathbf{x};\boldsymbol{\beta})\right]_{\hat{\boldsymbol{\beta}}}^{T}
\, \mathrm{Cov}(\hat{\boldsymbol{\beta}})
\,
\left[\nabla_{\boldsymbol{\beta}} S(\mathbf{x};\boldsymbol{\beta})\right]_{\hat{\boldsymbol{\beta}}} \\
&=&
\left[
S(\mathbf{x};\hat{\boldsymbol{\beta}})
\left(1-S(\mathbf{x};\hat{\boldsymbol{\beta}})\right)
\right]^2
\boldsymbol{\phi}(\mathbf{x})^{T}
\mathrm{Cov}(\hat{\boldsymbol{\beta}})
\boldsymbol{\phi}(\mathbf{x}).
\end{eqnarray}

We adopt a fiducial threshold of ${\rm [S/N]}_{\rm thresh}=12$, chosen to ensure consistency with previous CHIME/FRB population analyses \citep{2021ApJS..257...59C,ShinEtAl}. In addition, we also perform the fit at ${\rm [S/N]}_{\rm thresh}=8$ and ${\rm [S/N]}_{\rm thresh}=15$ to probe how the inferred selection function changes as the observation threshold is raised. The lower threshold corresponds to the lower limit for triggering by the search algorithm in L1 and includes a larger population of detections near the noise floor, while the higher threshold (${ \rm [S/N]_{thresh}}=15$) isolates a subset of events that are more separated from the noise. 

We fit the model using a reweighted least squares optimization algorithm from \citet{AllOfStatistics}, which we outline in Appendix~\ref{sec:IRLS}. This method, referred to as iteratively reweighted least squares (IRLS), is computationally efficient for logistic regression with our large population of injections and provides both the maximum likelihood estimates of the model coefficients and their associated covariance matrix. To restrict the selection function model to areas of parameter space physically probed by the injections, we implement a boolean K-nearest neighbors (KNN) mask in \texttt{scikit-learn} \citep{scikit-learn}. We set the threshold for the KNN mask arbitrarily at 12$\sigma$. We find that a fourth-degree polynomial expansion is sufficient to capture the selection probability surface without introducing unneccesary higher-order terms. We describe our model selection and validation procedure, including an introduction to the statistical metrics we use, in Appendix \ref{sec:model_selection}. The code used to construct and evaluate this selection function is publicly available at \url{https://github.com/CHIMEFRB/chimefrb-selection}. 

\begin{figure*}
    \centering
    \includegraphics[width=0.85\linewidth]{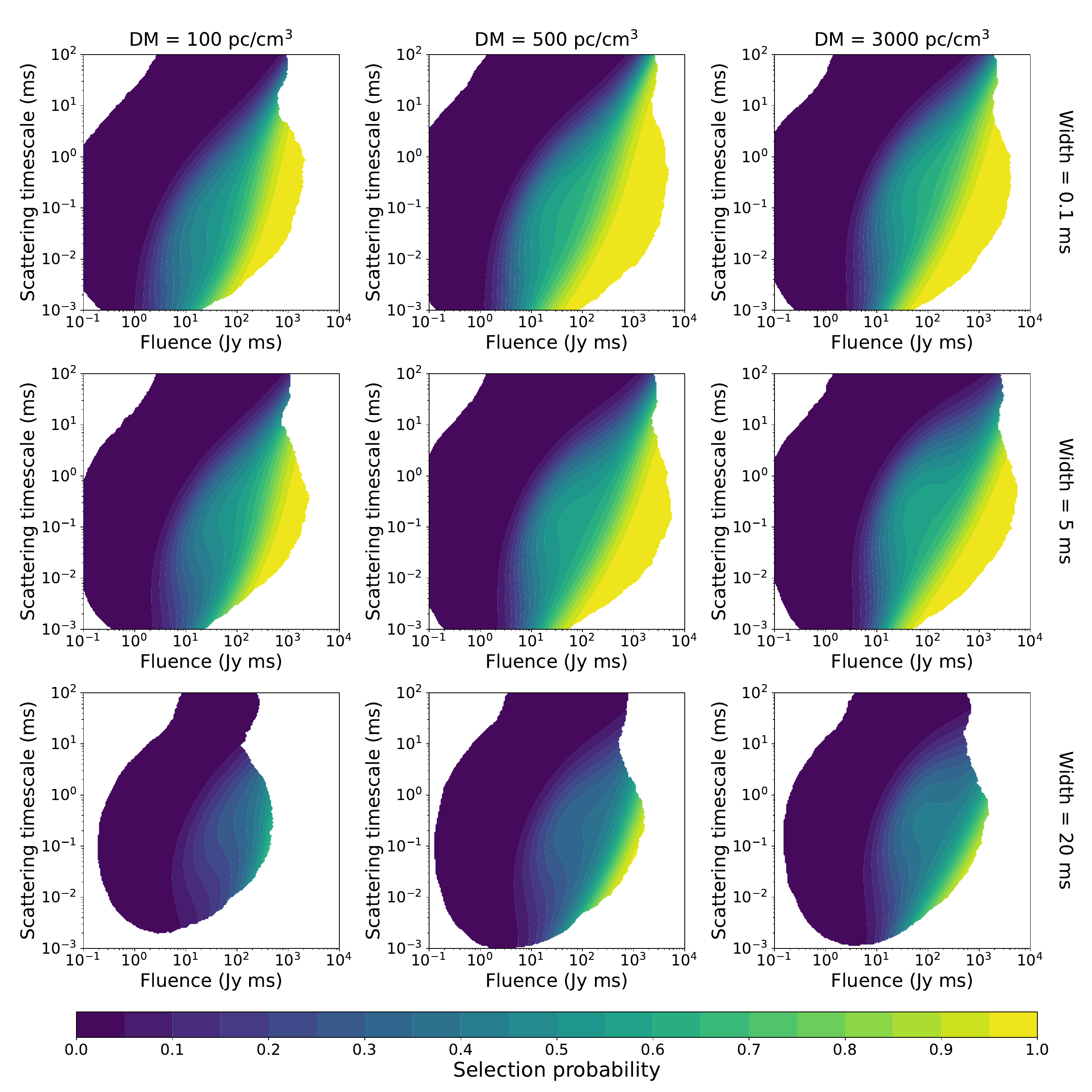}
    \caption{Detection probability from the selection function model evaluated on a grid of fluence and scattering timescale. Each panel shows $S(F,\tau\,|\,w,{\rm DM})$ at fixed intrinsic width $w$ (rows) and dispersion measure ${\rm DM}$ (columns). White areas correspond to masked regions where injections are not well-sampled. The color scale gives $S\in[0,1]$.}
    \label{fig:fluence_scattering_selection_probability_grid}
\end{figure*}

We provide a visualization of the selection function model in Figure~\ref{fig:fluence_scattering_selection_probability_grid}. Here we show slices of the selection probability surface across fluence and scattering timescale space at fixed intrinsic widths and DMs. The dominant trend is an increase in detection probability with fluence, but the location and shape of this selection surface depends strongly on scattering time and pulse width. At fixed DM and width, larger scattering times shift the detection boundary to higher fluence, while broader intrinsic widths substantially suppress the region of high selection probability. The presence of these correlations motivates the use of the full four-dimensional selection model, as the marginalized procedure applied in Section~\ref{sec:selection-correction-marginal} does not capture coupled changes in selection probability across $(F,\mathrm{DM},\tau,w)$.

\section{Refinement of the scattering distribution}
\label{sec:refinement-scattering-distribution}

The scattering timescale analysis presented in Section~\ref{sec:selection-correction-marginal} suggests an apparent downturn in the distribution of scattering times at large values. However, this interpretation is potentially affected by strong observational selection effects. In particular, the marginalized selection function for scattering times does not fully capture correlations between scattering and other burst properties that influence detectability. As highly scattered bursts are strongly suppressed by the detection pipeline, visualized in Figure~\ref{fig:width_dm_binned}, the selection-corrected distribution of scattering times presented therein may be significantly biased due to a highly correlated selection surface, as shown in Figure~\ref{fig:fluence_scattering_selection_probability_grid}.

To address this limitation, we apply the selection function model defined in Section~\ref{sec:selection-function-modeling}. The goal is to determine whether the apparent high-$\tau$ downturn reflects an intrinsic feature of the FRB population or instead arises from the observational selection effects. Motivated by the marginalized analysis, we parameterize the intrinsic scattering distribution using a model that reproduces the lognormal distribution inferred at low scattering times while allowing flexibility in the high-$\tau$ tail. This parameterization is written as follows:

\begin{equation}
p(\tau \mid \kappa, \tau_{\rm pivot}) \propto
\begin{cases}
\displaystyle
\frac{1}{\tau \theta_{\tau} \sqrt{2\pi}}
\exp\!\left[-\frac{\left(\ln \tau - \ln \tau_{\rm pivot}\right)^2}{2 \theta_{\tau}^2}\right],
& \tau_{\min} \le \tau \le \tau_{\rm pivot}, \\[1.2em]
\displaystyle
\frac{1}{\tau_{\rm pivot} \theta_{\tau} \sqrt{2\pi}}
\exp\!
\left(\frac{\tau}{\tau_{\rm pivot}}\right)^{\kappa-1},
& \tau_{\rm pivot} < \tau \le \tau_{\max}.
\end{cases}
\label{eq:tau_piecewise_beta}
\end{equation}
Here, $\theta_{\tau}=1.81$ is the logarithmic width of the fiducial scattering distribution (see Table \ref{tab:cat1_cat2_params}) and $\tau_{\rm pivot}$ is the peak of the lognormal half, which is kept as a free parameter. The model distribution therefore follows a lognormal below $\tau_{\rm pivot}$ and transitions to a power-law tail above this value. The slope of this tail is controlled by a single parameter $\kappa$ that determines whether the distribution turns downward, remains approximately flat, or increases toward larger scattering times. In this parameterization, $\kappa < 0$ corresponds to a declining tail, $\kappa = 0$ produces an approximately flat distribution in logarithmic space, and $\kappa > 0$ yields an upturn toward larger scattering times. Our objective is therefore to infer the values of $\kappa$ and $\tau_{\rm pivot}$ that best explain the observed catalog when accounting for the selection function. As the likelihood function is analytically complex once the full correlated selection function model and forward simulation are performed, we opt to employ a simulation-based inference (SBI) approach. Notably, SBI is commonly applied in astrophysics for population inference of transients \citep[e.g.,][]{2025ApJ...986...88S}, and our procedure is a useful worked example for further use of the selection function model presented in the previous section.

\begin{figure}[t]
    \centering
    \includegraphics[width=0.7\linewidth]{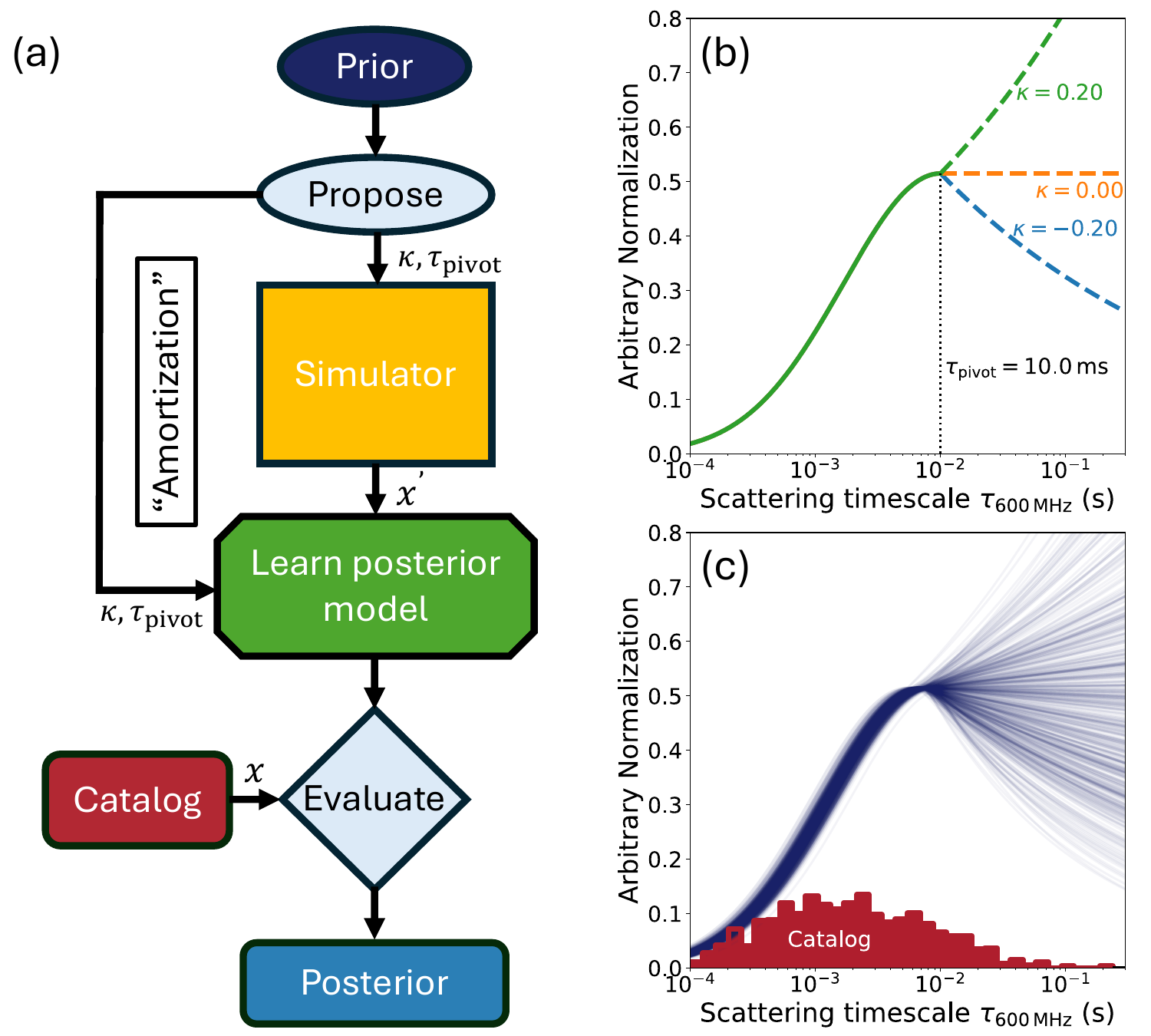}
    \caption{SBI framework used to constrain the scattering parameters $\kappa$ and $\tau_{\rm pivot}$. (a) Workflow of amortized posterior estimation, based on Figure 3f from \citet{FrontierOfSBI}. Parameters $\kappa$ and $\tau_{\rm pivot}$ are drawn from a uniform prior and passed through a forward simulator of scattering timescale detections to generate synthetic datasets $x^{\prime}$. These simulated parameter-data pairs are used as training data for a neural density estimator that learns an amortized model of the posterior distribution $p(\kappa,~\tau_{\rm pivot}~|~x)$. Once trained, the learned posterior model is then evaluated using the observed catalog data to obtain the inferred posterior distribution. (b)~Example models for several values of $\kappa$ demonstrating a downturn, plateau, and upturn in the distribution with $\tau_{\rm pivot}=10$ ms set at the fiducial value. (c) Random draws from the inferred posterior distribution, plotted as the range of model distributions consistent with the catalog data, plotted above the catalog scattering timescale histogram.}
    \label{fig:AmortizedInference}
\end{figure}

\subsection{Simulation-based inference of the scattering-tail slope}

We estimate the posterior distribution $p(\kappa,~\tau_{\rm pivot}\!\mid\!x)$ using neural posterior estimation (NPE), implemented with the \texttt{sbi} Python framework \citep{tejero-cantero2020sbi}. In this approach, the forward model of the FRB population and survey selection effects are treated as a simulator that generates synthetic observations for given model parameters. For each simulated realization, a value of $\kappa$ and $\tau_{\rm pivot}$ is first drawn from a uniform prior. Using this parameter vector, we then generate a simulated sample of FRB properties using the distributions for fluence, DM, and pulse width from the fiducial model and scattering times from the parameterized distribution in Equation~\ref{eq:tau_piecewise_beta}. For each parameter draw, we generate a weighted sample of FRB realizations and continue sampling until reaching a target effective sample size of $n_{\rm eff} \approx 10,\!000$. Similar to Equation~\ref{eqn:tau_neff}, we quantify the effective sample size using Kish’s design effect for weighted data \citep{kish1965survey},
\begin{equation}
    n_{\mathrm{eff}} = \frac{\left(\sum_i S(x_i) \right)^2}{\sum_i \left(S(x_i) \right)^2},
\end{equation}
where $S(x_i)$ are the detection probability for each simulated event that acts as a weighting for the event. This criterion ensures that each simulation has a comparable statistical precision despite variability in the weight distribution induced by the selection function. Each resulting mock catalog realization is summarized using a vector of quantiles of $\log \tau$ together with the mean and standard deviation of the distribution. This summary statistic provides a compact representation of the simulated scattering-time distribution while preserving information about the distribution point estimates. Pairs of simulated parameters and summary statistics $(\kappa,~ \tau_{\rm pivot}, ~x^\prime)$ are then used as training data for a neural density estimator \citep{FrontierOfSBI, pytorch} that learns an amortized approximation to the posterior distribution $p(\kappa, ~\tau_{\rm pivot}\!\mid\!x)$. Once training is complete, the observed catalog summary statistic vector is then provided to the trained network to obtain posterior samples for $\kappa$ and $\tau_{\rm pivot}$. We illustrate the inference framework in Figure~\ref{fig:AmortizedInference}.

\begin{figure}
    \centering
    \includegraphics[width=0.5\linewidth]{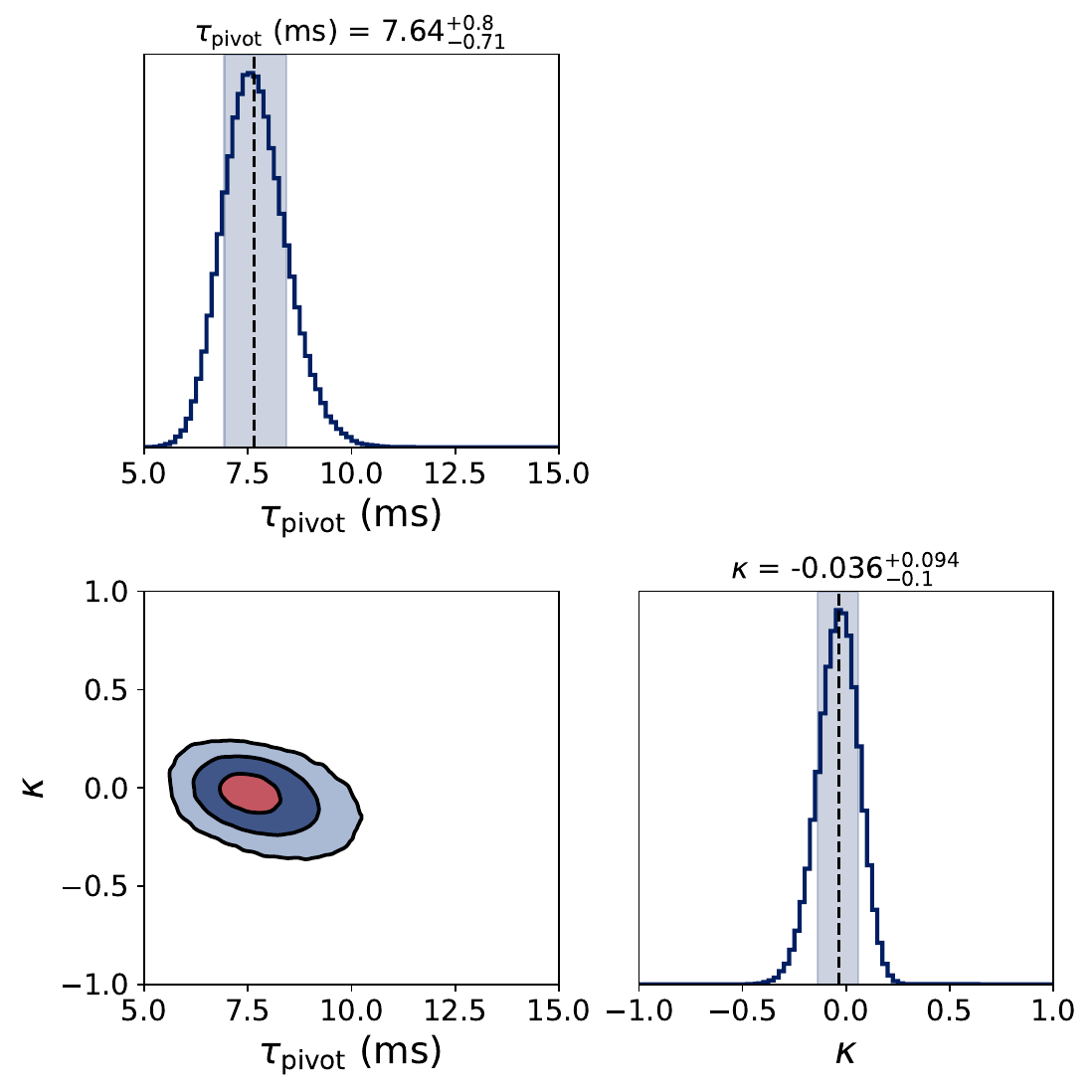}
    \caption{Corner plot of the estimated posterior distribution for the scattering model parameters $\tau_{\rm pivot}$ and $\kappa$. Contours enclose the central 68.3, 95.4, and 99.7 percent credible regions, respectively, while the diagonal panels show the marginalized distributions with median and 68\% credible intervals indicated.}
    \label{fig:sbi-corner-plot}
\end{figure}

The marginalized posterior for $\kappa$ is centered at a median of $\kappa\!=\!-0.036$ with a $95\%$ credible interval of $[-0.244, 0.143]$. The $\tau_{\rm pivot}$ marginalized posterior has a median of $\tau_{\rm pivot}=7.64$~ms with a $95\%$ credible interval of $[6.31, 9.31] $~ms. We show the posterior corner plot in Figure~\ref{fig:sbi-corner-plot}. These results indicate that the catalog is consistent with an approximately flat tail in logarithmic space at large scattering timescales. However, the posterior strongly disfavors strongly positive values of $\kappa$ that would imply a rapidly increasing abundance of highly scattered bursts, while moderately negative slopes remain allowed.  There also does not appear to be any strong correlation between $\tau_{\rm pivot}$ and $\kappa$. Random draws from the inferred posterior distribution demonstrate the range of scattering distributions consistent with the data, which exhibit a transition from the lognormal rise to a shallower tail, shown in Figure~\ref{fig:AmortizedInference}d. These results suggest that while the intense selection effect against high-scattering bursts primarily drives the drop-off in the observed distribution, the intrinsic scattering distribution is consistent with a uniform or slightly decreasing population at high scattering timescales.

\section{Discussion}
\label{sec:discussion}

The primary result of this work is the correction of survey selection effects out of the empirical fluence, DM, width, and scattering distributions for CHIME/FRB Catalog 2. Using a similar procedure as \citet{2021ApJS..257...59C} we fit a fiducial model to the selection-corrected marginalized distributions of these parameters. We then fit an logistic regression model for the selection probability across the parameter space of injections, as shown in Figure~\ref{fig:fluence_scattering_selection_probability_grid}, and apply this model to correct the scattering distribution in the presence of interactions between burst parameters. Here we discuss the implications of the scattering, width, and fluence distributions in the context of previous work, and provide guidance on future applications of the selection function model.

\subsection{Scattering timescale distribution}

\begin{figure*}[t]
\centering
\includegraphics[width=0.8\linewidth]{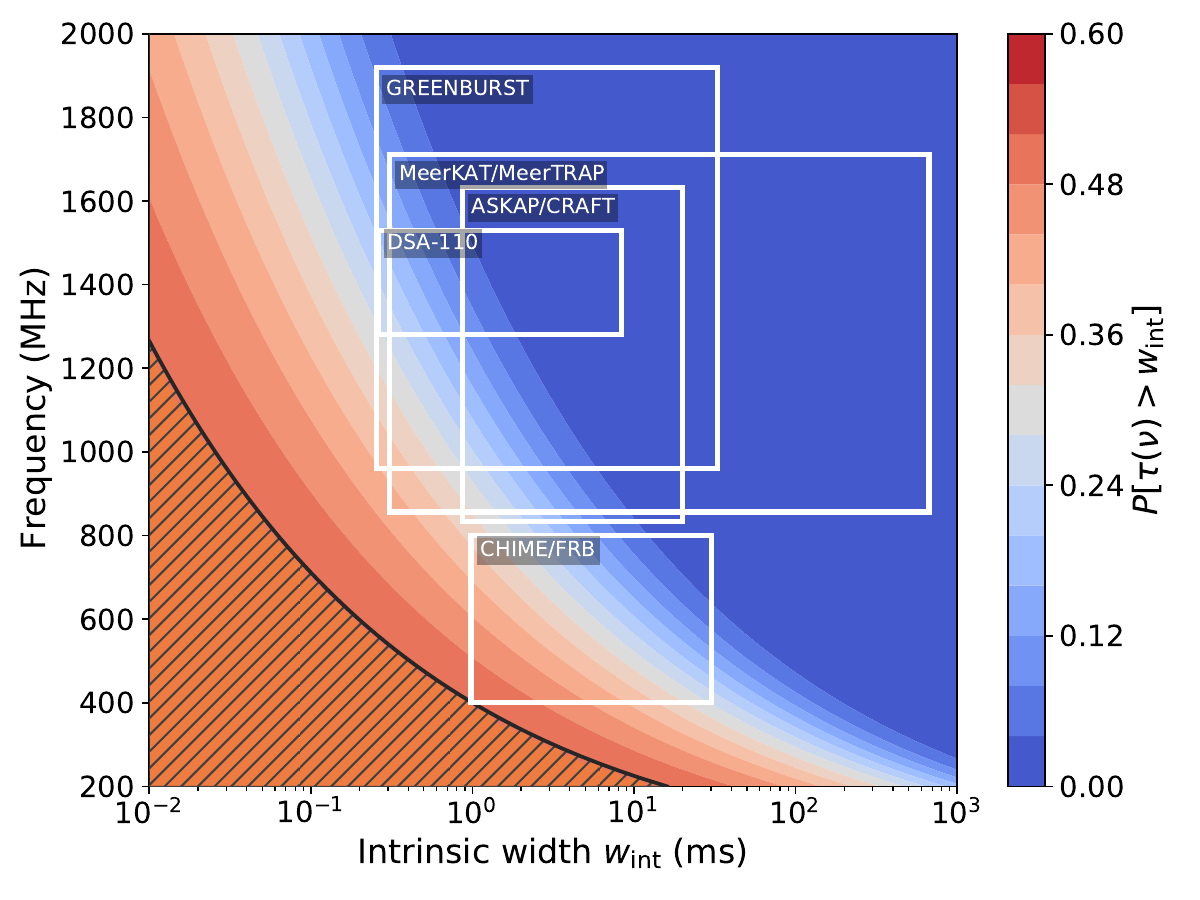}
\caption{Fraction of intrinsic FRB population with measured scattering timescale exceeding the intrinsic width, $P[\tau(\nu) > w_{\rm int}]$, computed using the Gaussian KDE in Figure~\ref{fig:scat-time-kernels}. Regions where this fraction approaches unity correspond to bursts with morphology that is dominated by scattering. Marked rectangles show the frequency coverage and the searched width range for select FRB surveys---including GREENBURST \citep{2026MNRAS.548ag665K}, MeerKAT/MeerTRAP \citep{2023MNRAS.524.4275J}, ASKAP/CRAFT \citep{2025PASA...42...36S}, DSA-110 \citep{2024ApJ...967...29L}, and CHIME/FRB \citep{CHIMEFRBSystemOverview}---extending from the time resolution to the maximum search width. Regions corresponding to scattering timescales below $\sim\!1$ ms at 400 MHz are hatched to reflect the uncertainty in the distribution shape below CHIME's time resolution, where scattering cannot be measured in Catalog 2.} 
\label{fig:fraction_scattered_above_width_2D_KDE}
\end{figure*}

We have constructed our analysis methods to leverage the statistical sample of Catalog~2 and improve constraints on the selection-corrected scattering distribution in a fully self-consistent manner. One benefit of our framework is the defining of the upper boundary of the recoverable scattering-time distribution in a way that is directly comparable to earlier work, while also constraining the nature of the intrinsic distribution at high values of scattering. Previous analyses by \citet{Pragya} and \citet{2021ApJS..257...59C} predicted the existence of a substantial population of highly scattered bursts that were largely undetectable to the intensity pipeline given the limited catalog size and injection coverage available at the time. The caveat, however, is that those studies tested for the existence of such a population, without the sensitivity to determine the structure within the distribution itself.

With the selection-corrected scattering distribution, we can compute the fraction of bursts whose scattering timescale exceeds their intrinsic width, $P[\tau(\nu) > w]$, as a function of observing frequency and intrinsic pulse width. This quantity measures the degree to which the observed burst morphology is dominated by propagation effects rather than the emission process itself, under the assumption of an intrinsically Gaussian temporal profile. Figure~\ref{fig:fraction_scattered_above_width_2D_KDE} visualizes this fraction across a broad region of frequency–width parameter space, with values approaching unity indicating bursts that are scattering-dominated. Overlaid rectangles denote the frequency coverage and searched width ranges of representative surveys, spanning from their native time resolution to the maximum width included in the search. In this representation, CHIME/FRB occupies a transition region between scattering- and morphology-dominated regimes, implying that its detected population includes a mixture of heavily over-scattered and negligibly scattered events. By contrast, surveys operating at higher frequencies probe regions where the fraction of over-scattered bursts is smaller.

\begin{figure*}[t]
    \centering
    \includegraphics[width=0.7\linewidth]{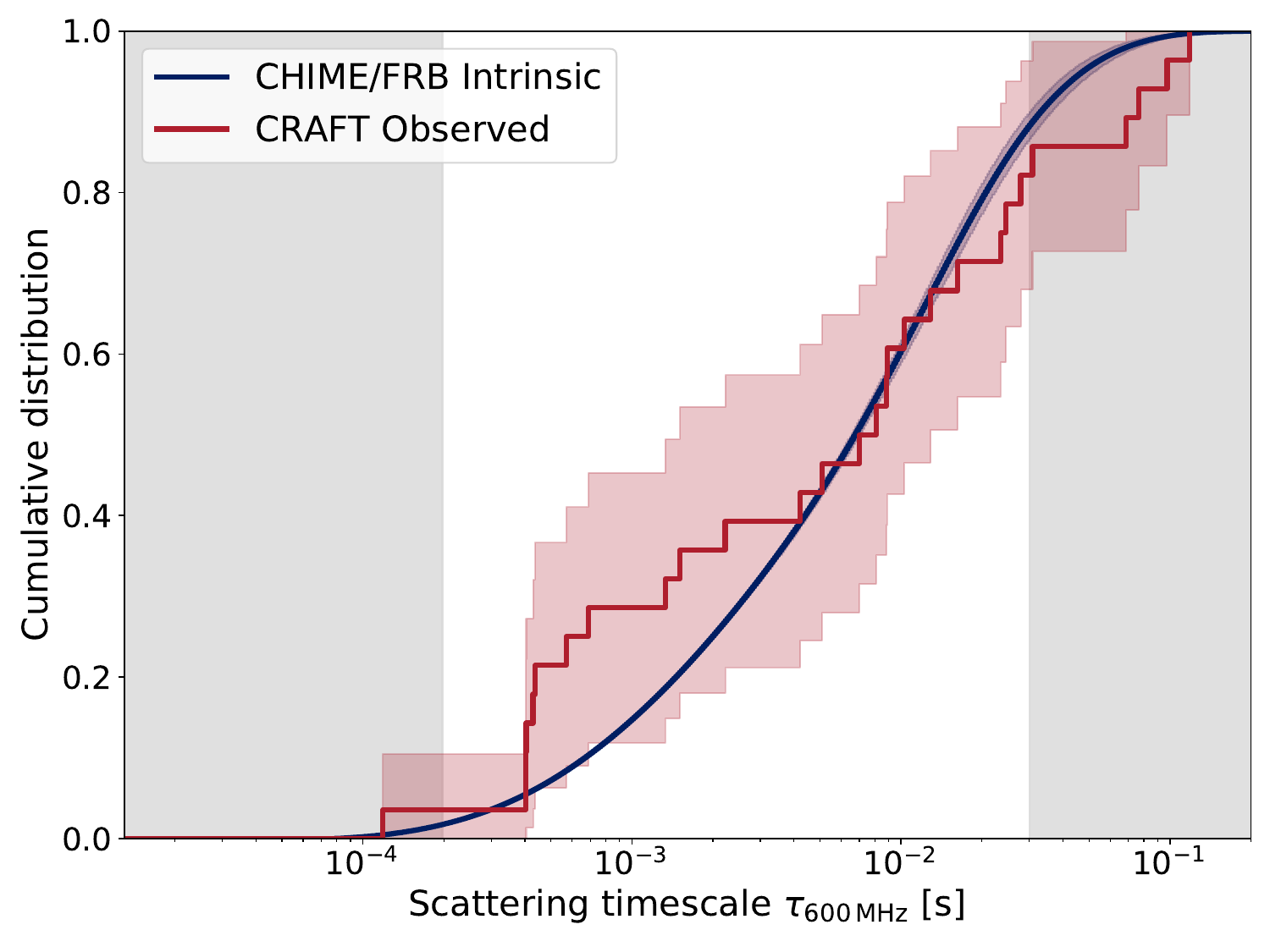}
    \caption{Scattering timescale CDFs for the CRAFT observed sample and inferred CHIME/FRB intrinsic population. Measurements scale the observed scattering time to a reference frequency of 600~MHz from $\tau_{\rm obs}$ by \citet{2025PASA...42..133S} assuming a power-law scattering index of $-4$. (Red) CDF of the observed CRAFT scattering timescales with 95\% pointwise confidence interval, representing the finite-sample uncertainty in the KDE distribution. (Blue) CDF of the CHIME/FRB Catalog~2 intrinsic scattering timescale KDE with $95\%$ uncertainty interval from a bootstrap method. The light gray vertical region indicates $\tau < 1$ ms at 400 MHz, where no scattering can be measured by CHIME/FRB, and $\tau \ge 30$~ms, where the observed distribution is increasingly uncertain.}
    \label{fig:craft_compare}
\end{figure*}

Figure~\ref{fig:craft_compare} compares the scattering timescale CDF from the sample observed by the CRAFT survey \citep{2025PASA...42..133S} to the inferred intrinsic CHIME/FRB population, with both distributions scaled to a reference frequency of 600~MHz assuming \(\tau \propto \nu^{-4}\). The CRAFT distribution is broadly consistent with the selection-corrected CHIME/FRB distribution. This comparison differs from that of \citet{2025PASA...42..133S}, who compared the observed CRAFT and CHIME/FRB distributions at 1000~MHz and 600~MHz, respectively, and suggested that the apparent high-\(\tau\) dropoff could reflect a common selection effect against highly scattered bursts.

With both samples referenced to 600~MHz, the consistency between the observed CRAFT distribution and the inferred intrinsic CHIME/FRB distribution suggests that CHIME/FRB is not missing a large population of scattered bursts relative to the population probed by CRAFT. We reason this as CRAFT observes at higher frequencies, where bursts with large scattering times at CHIME frequencies should be less temporally broadened and therefore easier to detect. The absence of a substantially heavier high-\(\tau\) tail in CRAFT therefore supports the conclusion that the intrinsic scattering-timescale distribution does not rise strongly above \(\tau \sim 10\)~ms. As catalog events along sightlines for which Galactic scattering is expected to be measurable are excised from the sample, the presence of this behavior is unlikely to be an artifact of Milky Way scattering and provides some of the strongest evidence for an extragalactic population of highly scattered bursts.

We also emphasize, however, that at lower scattering timescales the comparison remains limited by the survey time resolution. CHIME/FRB cannot measure any scattering value below approximately 1~ms at 400~MHz in Catalog~2, and larger values may also not be measured due to domination by the intrinsic pulse width (see Figure~\ref{fig:cat2-scatULs-withfluence}). CRAFT is similarly insensitive due to analogous effects at low scattering timescales relative to its higher frequency range and bandwidth. However, as FRBs may be intrinsically narrower at higher radio frequencies, it may be easier to resolve subtle scattering tails \citep{2018ApJ...863....2G}.  While the current data disfavor a strongly rising high-\(\tau\) tail, we cannot strongly distinguish between a distribution that is approximately uniform in \(\log \tau\) and one with additional structure at the low-\(\tau\) end.

\subsection{Intrinsic width distribution}
Both the observed and selection-corrected width distributions show an accumulation of bursts at widths below the CHIME/FRB time resolution of 0.983~ms (see Fig.~\ref{fig:width_dm_binned}). This feature could suggest an excess of sub-millisecond bursts, but may also be due to systematic effects in the width estimation procedure. In particular, the \texttt{fitburst} modeling framework \citep{Fitburst}, which fits parametric templates to burst dynamic spectra, may bias width estimates toward artificially small values when the underlying burst morphology deviates from the assumed model. High time-resolution observations of FRBs, including follow-up of repeating sources and population studies with CHIME/FRB baseband data, have revealed complex non-gaussian temporal structures, often consisting of multiple unresolved sub-bursts and fine temporal features \citep{2023MNRAS.526.2039H, 2025ApJ...979..160S, 2025ApJ...992..206C}. Individual burst components may be intrinsically asymmetric, with non-gaussian temporal profiles produced by the emission process itself rather than propagation effects. When these intrinsically structured signals are modeled with simplified functional forms, the inferred widths may collapse to values below the instrumental resolution.

\subsection{Fluence distribution}

In the Catalog~1 analysis, the joint distribution of fluence and DM was modeled using a parametric form, motivated by the possibility of a DM-dependent fluence slope \citep{2021ApJS..257...59C}. That work employed both maximum-likelihood estimation and Markov Chain Monte Carlo (MCMC) sampling of the posterior. The resulting constraints were $\alpha = -1.32 \pm 0.11$ from maximum likelihood and $\alpha = -1.41 \pm 0.11$ from the MCMC analysis. These estimates are mutually consistent within $1\sigma$, though the MCMC result provides a more complete characterization of parameter uncertainties and correlations. Notably, this parameter estimate is also consistent with the $\alpha = -1.5$ expectation for a non-evolving population in Euclidean space. In contrast, this work adopts the maximum-likelihood framework throughout. The resulting constraint $\alpha = -1.07 \pm 0.02$ is shallower, however, than both the maximum-likelihood and MCMC fit found by \citet{2021ApJS..257...59C}, arising from the absence of an explicit joint treatment of fluence and DM and the exclusion of bursts with unmeasured scattering timescales. The cosmological evolution of FRB energetics will be addressed in a concurrent work by N. Jain et al. (in prep.) with an analysis following \citet{ShinEtAl}, incorporating an explicit treatment of redshift evolution and volumetric rate mixture estimates. As such, we defer a more optimal treatment of the distribution of FRB energetics to that work.

\subsection{Future applications of the selection function}

The selection function presented in this work is directly relevant to the interpretation of CHIME/FRB intensity data products and provides a framework for extending population inference to baseband data and CHIME/FRB Outrigger localizations \citep{Basecat1, OutriggerOverview, AdamKKOCatalog, RBFLOAT, 2024NatAs...8.1429C, VishwangiOutrigger}. CHIME/FRB saves raw voltage baseband data for only the highest S/N subset of the FRBs for which we capture total-intensity data, typically those with searched $\mathrm{S/N} > 12$ \citep{BasebandPipeline}. These baseband captures preserve raw complex voltages at microsecond time resolution, enabling coherent dedispersion, improved spectro-temporal modeling with \texttt{fitburst}, and more precise localization \citep{Basecat1}. Thus, the selection function may describe the probability that a burst with intrinsic properties $(F,\tau,w,\mathrm{DM})$ produces the intensity trigger required for downstream baseband capture and localization.

The triggering of baseband data introduces an additional hierarchical layer to the CHIME/FRB detection process. In current operations, only a subset of intensity triggers can trigger a baseband capture, and even among these, archiving is contingent on the availability of the baseband capture system. Importantly, the latter does not share identical uptime with the real-time intensity pipeline, leading to an effective selection that is not purely a function of S/N. Accounting for this is further complicated by time-variable changes in system performance or external observing environment are not currently well-modeled. Quantifying the relative uptime and overlap between the intensity and baseband systems is therefore necessary for a complete accounting of the baseband selection function; this characterization will be addressed in a forthcoming publication. A further complication arises from the fact that baseband triggering and recording are operationally distinct from the CHIME/FRB Outrigger backends. The CHIME/FRB Outrigger system, which consists of geographically separated stations across North America, receives real-time triggers from the CHIME core site and independently dumps baseband data for offline analysis \citep{2021AJ....161...81L, OutriggerOverview}. The selection function developed here is explicitly limited in scope to the probability of triggering intensity data at the CHIME core and does not model whether baseband dumps are successfully recorded at the core or at Outrigger stations. While these triggering layers are expected to be correlated, as they depend on the same underlying intensity detection, their joint selection behavior is not captured in the present framework.

We emphasize that CHIME/FRB fluence measurements derived from intensity data should be interpreted as lower limits, owing to uncertainty in burst localization within the primary beam and the resulting ambiguity in beam response correction \citep{2023AJ....166..138A}. One advantage of a parametric selection function model is that debiasing population-level distributions under these conditions is straightforward. In particular, the Horvitz-Thompson estimator \citep[see][]{HorvitzThompsonEstimator, 2018arXiv180404255Z} provides a natural and computationally trivial means of correcting for selection effects by weighting each detected burst by the inverse of its detection probability under the selection function. Looking ahead, the production of a large, systematically characterized catalog of baseband-localized events will enable substantially more precise population studies. Baseband data provide fluence and width measurements at microsecond precision with well-constrained beam responses, allowing these events to be incorporated directly into population synthesis models with minimal additional uncertainty. Once a baseband catalog corresponding to the intensity catalog is released, inverse probability weighting using the selection function model presented here will be straightforward to implement. However, this will require an explicit characterization of the ``baseband selection function'' as discussed above. Work toward modeling this extended selection hierarchy is currently underway, and the statistical framework applied in this paper provides a natural foundation for such future analyzes.

\section{Conclusion}
\label{sec:conclusion}
In this work, we have presented a population analysis of FRB observables in CHIME/FRB Catalog~2 using nearly an order of magnitude more synthetic bursts injected into the instrument's live search pipeline compared to previous works \citep{2021ApJS..257...59C, Merryfield}. These injections enable an empirical determination of the survey's response and provide the basis for both a revised fiducial model of the intrinsic FRB observable distributions and an explicit, multidimensional model of the CHIME/FRB selection function. Using the injection population, we first fit a fiducial model for the intrinsic distributions of fluence, DM, intrinsic width, and scattering timescale following the framework originally presented in \citet{2021ApJS..257...59C}. We find that the Catalog~2 data broadly support the qualitative picture established in earlier work, while yielding tighter constraints on the corresponding population parameter estimates. 

We then constructed a logistic regression model for the CHIME/FRB selection function in the four-dimensional space of fluence, DM, intrinsic width, and scattering timescale, of which the latter three are directly measured observable quantities for CHIME/FRB. This model provides an explicit parameterization of the CHIME/FRB detection probability, avoiding the assumption of separability adopted in earlier treatments. We showed that a fourth-order polynomial expansion is sufficient to capture the dominant structure of the selection surface.

The resulting selection model is useful both as a practical tool for debiasing population statistics and as a forward model for generating realistic mock detected samples. Applying this model, we refine the estimate of the scattering distribution to more robustly probe the behavior of the distribution at high scattering times. While the marginalized selection-correction estimates suggest an apparent downturn in the distribution above a characteristic peak at 13.1~ms (referenced to 600~MHz), our treatment suggests that this may be driven by the severe selection bias in this regime. Using simulation-based inference, we constrained the slope of the high-$\tau$ tail and found that the catalog data are consistent with an approximately flat or mildly declining distribution above the characteristic peak, though a continued upturn in the distribution cannot be ruled out. We further quantified the range over which the scattering distribution is meaningfully constrained, finding that the Catalog~2 data remain informative to larger scattering timescales than was possible in Catalog~1, conservatively raising the well-constrained scattering domain from 10~ms to around 30~ms at 600 MHz.

This work emphasizes that rigorous FRB population inference requires an equally rigorous treatment of survey selection effects. In the near term, it enables improved correction of population statistics from intensity data products and more physically interpretable comparisons with other FRB surveys. In the longer term, it provides a natural basis for extending selection modeling to other CHIME/FRB data products, including baseband data and Outrigger localizations, where additional layers of detectability and instrumental uptime should be accounted for. As FRB samples continue to mature, selection function modeling will become increasingly central to transforming survey catalogs into robust constraints on the intrinsic burst population and its underlying astrophysics. The methods presented here are intended as a step toward that goal: a survey-calibrated, explicitly modeled, and practically usable description of the CHIME/FRB selection function that can support future studies of FRB populations.


\section{Acknowledgments}

We acknowledge that CHIME is located on the traditional, ancestral, and unceded territory of the Syilx/Okanagan people. We are grateful to the staff of the Dominion Radio Astrophysical Observatory, which is operated by the National Research Council of Canada. CHIME operations are funded by a grant from the NSERC Alliance Program and by support from McGill University, University of British Columbia, and University of Toronto. CHIME was funded by a grant from the Canada Foundation for Innovation (CFI) 2012 Leading Edge Fund (Project 31170) and by contributions from the provinces of British Columbia, Québec and Ontario. The CHIME/FRB Project was funded by a grant from the CFI 2015 Innovation Fund (Project 33213) and by contributions from the provinces of British Columbia and Québec, and by the Dunlap Institute for Astronomy and Astrophysics at the University of Toronto. Additional support was provided by the Canadian Institute for Advanced Research (CIFAR), the Trottier Space Institute at McGill University, and the University of British Columbia. The CHIME/FRB baseband recording system is funded in part by a CFI John R. Evans Leaders Fund award to IHS.

The AstroFlash research group at McGill University, University of Amsterdam, ASTRON, and JIVE is supported by: a Canada Excellence Research Chair in Transient Astrophysics (CERC-2022-00009); an Advanced Grant from the European Research Council (ERC) under the European Union’s Horizon 2020 research and innovation programme (`EuroFlash’; Grant agreement No. 101098079); an NWO-Vici grant (`AstroFlash’; VI.C.192.045); an NSERC Discovery Grant (RGPIN-2025-06681); an ERC Starting Grant (`EnviroFlash’; Grant agreement No. 101223057); and an NWO-Veni grant (VI.Veni.222.295).

We thank Sujay Mate for helpful comments on the CHIME/Slow commissioning and science observing timeline.
K.T.M. is supported by an FRQNT Master's Research Scholarship.
M.W.S. is a Fonds de recherche du Qu\'ebec -- Nature et technologies (FRQNT) postdoctoral fellow and acknowledges support from the Trottier Space Institute Fellowship program.
R.V.C. is supported by the NSERC of Canada Discovery Grant RGPIN-2024-04506.
S.P.E.T. is a Fonds de recherche du Qu\'ebec -- Nature et technologies (FRQNT) doctoral fellow.
G.M.E. acknowledges support from NSERC via the Discovery Grant program (RGPIN-2020-04554).
D.C.S. is supported by an NSERC Discovery Grant (RGPIN-2021-03985).
P.S. acknowledges the support of an NSERC Discovery Grant (RGPIN-2024-06266).
V.M.K. holds the Lorne Trottier Chair in Astrophysics \& Cosmology and a Distinguished James McGill Professorship, and receives support from an NSERC Discovery Grant (RGPIN 228738-13).
D.B. is supported by an NSERC Discovery Grant.
K.W.M. is supported by NSF Grant No.~2510771.
A.B.P. acknowledges support by NASA through the NASA Hubble Fellowship grant HST-HF2-51584.001-A, awarded by the Space Telescope Science Institute, which is operated by the Association of Universities for Research in Astronomy, Inc., under NASA contract NAS5-26555.
A.B.P. also acknowledges prior support from a Banting Fellowship, a McGill Space Institute~(MSI) Fellowship, and a Fonds de recherche du Qu\'ebec -- Nature et technologies~(FRQNT) Postdoctoral Fellowship.
A.P. is a Trottier Space Institute Postdoctoral Fellow.
F.A.D. is funded by the Government of Canada / financ\'e par le gouvernement du Canada.
S.S.P. is supported by the National Science Foundation under Grant AST-2407399.
A.P.C. is a Canadian SKA Scientist and is funded by the Government of Canada / est financ\'e par le gouvernement du Canada.
C.L. acknowledges support from the Miller Institute for Basic Research at UC Berkeley.
A.M.C. is a Banting Postdoctoral Researcher.

The author also gratefully acknowledges the Montréal Canadiens’ 2026 Stanley Cup playoff run, during which much of this manuscript was developed and from which the color schemes of several figures drew inspiration.
%





\appendix

\section{Iteratively Reweighted Least Squares Algorithm}
\label{sec:IRLS}
We fit our selection function parameterization \( S(x | \boldsymbol{\beta}) \) using a form of the \textit{Iteratively Reweighted Least Squares} (IRLS) algorithm, as outlined in Section X of \citet{AllOfStatistics}. Let \( \mathbf{X} \) denote the polynomial design matrix with components \( x_{ij} \) corresponding to the \( i \)th predictor for the \( j \)th injection. At each iteration \( s \), we update the coefficient vector \( \boldsymbol{\beta}^{(s)} \) by reweighting the logit-transformed response.

Starting from the logistic model for the selection function,
\begin{equation}
S(x_i \mid \boldsymbol{\beta}^{(s)}) = \frac{1}{1 + \exp\left( - \sum_j \beta_j^{(s)} x_{ij} \right)},
\end{equation}
we construct a diagonal weight matrix \( W^{(s)} \) and adjusted response vector \( Z^{(s)} \):
\begin{equation}
W^{(s)} = \mathrm{diag} \left( S(x_i \mid \boldsymbol{\beta}^{(s)}) \left[ 1 - S(x_i \mid \boldsymbol{\beta}^{(s)}) \right] \right),
\end{equation}
\begin{equation}
Z_i^{(s)} = \mathrm{logit}\left(S(x_i \mid \boldsymbol{\beta}^{(s)})\right) + \frac{y_i - S(x_i \mid \boldsymbol{\beta}^{(s)})}{W_i^{(s)}}.
\end{equation}
This transforms the problem into a weighted linear regression of \( Z^{(s)} \) on \( \mathbf{X} \) at each step:
\begin{equation}
\boldsymbol{\beta}^{(s+1)} = \left( \mathbf{X}^\top W^{(s)} \mathbf{X} + \lambda \mathbf{I} \right)^{-1} \mathbf{X}^\top W^{(s)} Z^{(s)},
\end{equation}
where \( \lambda \) is a small regularization parameter and \( \mathbf{I} \) is the identity. The procedure is iterated until convergence, and within 500 iterations we do not observe any signs indicating against convergence.

\section{Model Selection and Validation}
\label{sec:model_selection}

\begin{figure*}
    \centering
    \includegraphics[width=\linewidth]{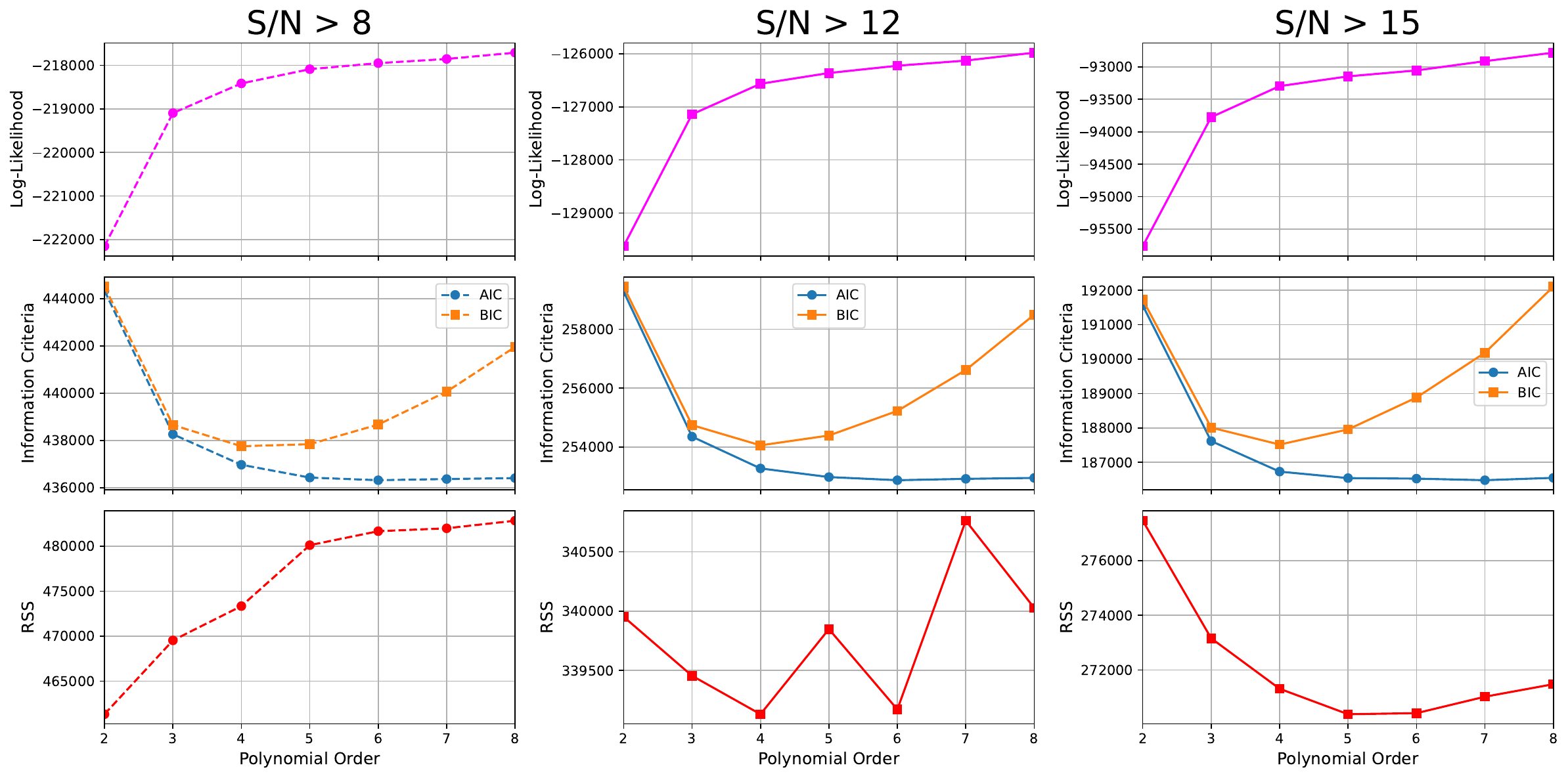}
    \caption{Relative model comparison summary statistics for logistic regression fits of the injected population with ${\rm [S/N]}_{\rm thresh} =\{ 8, 12, 15\}$ as a function of increasing polynomial order. Plotted statistics include the Log-Likelihood, Akaike Information Criterion (AIC), Bayesian Information Criterion (BIC), Pearson residual sum of squares (RSS). These diagnostics are intended for comparing the relative performance of models across different orders, in compliment to the absolute goodness of fit statistics calculated in Table~\ref{tab:model-selection}.}
    \label{fig:model-selection}
\end{figure*}

To determine the optimal functional form for our selection function parameterization, we consider a series of binary logistic regression models with increasing polynomial complexity. Each model of polynomial order $n$ is defined by a design matrix $\mathbf{X} \in \mathbb{R}^{N \times P_n}$, where $N$ is the number of injections and $P_n$ is the number of polynomial basis terms (including interaction terms) up to total degree $n$ in the four predictor variables: log-transformed fluence $F$, scattering time $\tau_{\rm scat}$, pulse width $w$, and dispersion measure DM. We construct design matrices for each model up to polynomial order $n \in \{1, 2, \ldots, 7\}$. The total number of terms $P_n$ grows combinatorially with $n$ and the number of predictors $d = 4$, following $P_n = \binom{n + d}{d}.$ This formulation allows us to construct increasingly flexible logistic regression models with $\hat{\boldsymbol{\beta}} \in \mathbb{R}^{P_n}$ as the vector of regression coefficients estimated from the data. For model selection diagnostics we calculate the log-likelihood, relevant information criteria, and the sum of squares for Pearson residuals (RSS); these are shown in Figure \ref{fig:model-selection}. We also calculate McFadden’s pseudo-$R^2$ statistic,  mean Brier score across 5 folds, and Receiver Operating Characteristic Area Under Curve (ROC AUC), which are presented in Table \ref{tab:model-selection}. These each serve as metrics for assessing the predictive performance of binary classification models; we introduce them briefly in Appendix~\ref{sec:model_selection}.

In general, we favor models that balance predictive performance with parsimony, avoiding unnecessary complexity once improvements in fit become marginal. Figure~\ref{fig:model-selection} summarizes the relative model-comparison diagnostics for each S/N threshold as a function of polynomial order. For all three thresholds, the log-likelihood increases between second and fourth order, suggesting that low-order models are underfitting the structure of the selection surface. Beyond approximately fourth order, however, the log-likelihood curves flatten, suggesting that additional polynomial terms yield diminishing improvements in fit. The information criteria show similar behavior but with clearer penalties for complexity. In all cases, differences among models of order $\gtrsim4$ are small compared to the improvements observed between low-order models, indicating that most of the explanatory structure is captured by relatively modest polynomial complexity.

The Pearson RSS provides another measure of model misfit. For S/N~$>8$, the RSS decreases immediately from order 2. For S/N~$>12$, the RSS reaches its minimum near order 4 and fluctuates for higher orders. The S/N~$>15$ fit shows a smoother pattern, likely due to the increasingly stringent threshold for detection, with a minimum around fifth order. These trends indicate that the majority reduction of the residuals occurs by fourth order, and that additional terms do not substantially improving the fit.

While Figure~\ref{fig:model-selection} evaluates relative goodness-of-fit and model complexity, Table~\ref{tab:r2_brier_summary} assesses predictive performance using cross-validation metrics. These display behavior consistent with the selection diagnostics. The Brier score \citep{Brier1950} decreases rapidly from first to third order and then changes only at the $\sim10^{-4}$ level for higher orders, while the ROC AUC \citep[see][]{2002QJRMS.128.2145M} increases from low-order models and quickly saturates. Likewise, McFadden's pseudo-$R^2$ \citep{McFadden_1979} rises steeply at low order but exhibits only marginal gains beyond fourth or fifth order for all S/N thresholds. $R^2_{\mathrm{McF}}$ provides a measure of model fit based on the likelihood ratio of nested models, and values in the range $0.2$–$0.4$ canonically indicate excellent predictive performance for discrete choice models \citep[see][]{McFadden_1979}. All models of order $n \geq 2$ lie within or above this range for all S/N thresholds, indicating that even relatively low-order models already provide a good description of the data. However, the increase in $R^2_{\mathrm{McF}}$ beyond fourth order is small (e.g., $\Delta R^2_{\mathrm{McF}} \sim 0.005$ between $n=4$ and $n=6$ for S/N~$>12$), corresponding to a fractional change of $\sim1\%$, which evidences that higher-order terms yield only marginal improvements in explanatory power despite substantially increasing model complexity. Furthermore, increasing the S/N threshold yields uniformly larger $R^2_{\mathrm{McF}}$ values and ROC AUC scores, as higher thresholds suppress false positive detections and thereby improve the separability between detected and non-detected injections. The parity between model selection and cross-validation diagnostics indicates that models of order $\gtrsim4$ provide nearly equivalent performance, and that this is consistent across all S/N thresholds. Because higher-order models introduce substantially more parameters without delivering commensurate gains in information criteria, residual reduction, or predictive accuracy, they offer little practical advantage and risk fitting noise in the injected population. We therefore adopt the fourth-order polynomial model as our fiducial parameterization, as it lies within the plateau region of all performance metrics while maintaining a relatively parsimonious functional form.
\begin{deluxetable*}{cc|ccc|ccc|ccc} 
\label{tab:model-selection}
\tablecaption{McFadden's pseudo-$R^2$, mean Brier score (BS) across 5 folds, and Receiver Operating Characteristic Area Under Curve (ROC AUC) for models of increasing polynomial order, evaluated on the S/N~$> 8$, S/N~$> 12$, and S/N~$> 15$ injection populations.\label{tab:r2_brier_summary}}
\tablewidth{0pt}
\setlength{\tabcolsep}{5pt}
\tablehead{
\colhead{Order} &
\colhead{$P_n$} &
\multicolumn{3}{c}{S/N~$ > 8$} &
\multicolumn{3}{c}{S/N~$ > 12$} &
\multicolumn{3}{c}{S/N~$ > 15$} \\
\colhead{} &
\colhead{} &
\colhead{$R^2_{\mathrm{McF}}$} &
\colhead{$\langle \mathrm{BS}\rangle$} &
\colhead{ROC AUC} &
\colhead{$R^2_{\mathrm{McF}}$} &
\colhead{$\langle \mathrm{BS}\rangle$} &
\colhead{ROC (AUC)} &
\colhead{$R^2_{\mathrm{McF}}$} &
\colhead{$\langle \mathrm{BS}\rangle$} &
\colhead{ROC AUC}
}
\startdata
1 & 5   & 0.2153 & 0.14002 & 0.8251 & 0.2902 & 0.07872 & 0.8882 & 0.3220 & 0.05723 & 0.9120 \\
2 & 16  & 0.2633 & 0.1330  & 0.8405 & 0.3459 & 0.07459 & 0.8999 & 0.3803 & 0.05415 & 0.9218 \\
3 & 35  & 0.2734 & 0.1318  & 0.8420 & 0.3584 & 0.07366 & 0.9013 & 0.3932 & 0.05341 & 0.9232 \\
4 & 70  & 0.2757 & 0.13157 & 0.8425 & 0.3614 & 0.07350 & 0.9016 & 0.3963 & 0.05328 & 0.9234 \\
5 & 126 & 0.2768 & 0.13146 & 0.8426 & 0.3624 & 0.07344 & 0.9017 & 0.3973 & 0.05325 & 0.9235 \\
6 & 210 & 0.2773 & 0.13145 & 0.8426 & 0.3631 & 0.07343 & 0.9017 & 0.3979 & 0.05327 & 0.9234 \\
7 & 330 & 0.2776 & 0.13149 & 0.8425 & 0.3636 & 0.07349 & 0.9015 & 0.3988 & 0.05330 & 0.9233 \\
\enddata
\end{deluxetable*}

\subsection{Log-likelihood}

As the selection function $S(\mathbf{x})$ represents the probability that a burst with properties $\mathbf{x}$ is detected by CHIME/FRB, we model the likelihood of the injection outcomes using the Bernoulli distribution. Each injection in the dataset is either a detection ($y_i = 1$) or a non-detection ($y_i = 0$), and the probability of observing each outcome is given by the predicted selection probability $S(\mathbf{x}_i)$ from the logistic model.

The likelihood function for the full sample of $N$ injections is then given by the product:
\[
\mathcal{L}(\hat{\boldsymbol{\beta}}) = \prod_{i=1}^{N} S(\mathbf{x}_i)^{y_i} \left[1 - S(\mathbf{x}_i)\right]^{1 - y_i},
\]
where $S(\mathbf{x}_i)$ is the predicted selection probability for injection $i$ and $\hat{\boldsymbol{\beta}}$ is the vector of converged regression coefficients for a given polynomial order model (see Appendix \ref{sec:IRLS}).

Taking the log-likelihood,
\[
\log \mathcal{L}(\hat{\boldsymbol{\beta}}) = \sum_{i=1}^{N} \left[ y_i \log S(\mathbf{x}_i) + (1 - y_i) \log (1 - S(\mathbf{x}_i)) \right].
\]
We maximize this quantity with respect to the regression coefficients $\boldsymbol{\beta}$ by using the fitting algorithm described in Appendix \ref{sec:IRLS}. The resulting log-likelihood value serves as a relative measure of model fit across the parameter space and is requisite for calculating various information criteria necessary to compare models of varying polynomial complexity.

\subsection{Information Criteria}
To compare logistic regression models of varying polynomial order, we compute two widely used information criteria: the Akaike Information Criterion (AIC) and the Bayesian Information Criterion (BIC). Both quantities balance model fit against model complexity, penalizing overparameterized models to avoid overfitting. They are defined as:
\[
\mathrm{AIC} = 2k - 2 \log \hat{\mathcal{L}},
\qquad
\mathrm{BIC} = k \log N - 2 \log \mathcal{\hat{\mathcal{L}}},
\]
where $k$ is the number of free parameters in the model (equal to the number of terms in the design matrix $P_n$), $N$ is the number of injections, and $\log \hat{\mathcal{L}}$ is the maximized log-likelihood of the model with estimated regression coefficients. Lower values of AIC and BIC indicate a more parsimonious model with better support from the data, also interpreted as minimizing the loss of information from the raw data to the model.


\subsection{Likelihood Ratio}

In addition to the likelihood-based information criteria, we compute McFadden’s pseudo-$R^2$ as a measure of overall model fit. This statistic is defined as
\[
R^2_{\text{McF}} = 1 - \frac{\log L_{\text{model}}}{\log L_{\text{null}}},
\]
where $\log L_{\text{model}}$ is the log-likelihood of the fitted model and $\log L_{\text{null}}$ is that of a null model containing only the $\hat{\beta}_0$ intercept term \citep{McFadden_1979}.
It is important to note that $R^2_{\rm McF}$ is {\it not} directly analagous to the coefficient of determination used in linear regression, which quantifies explained variance via the squared Pearson correlation and does not rely on parameter estimation {\it a priori}. However the pseudo-$R^2$ statistic can still be interpreted as a goodness of fit measure, as it quantifies the relative improvement in model fit over the null model, conditioned on the parameter estimates $\mathbf{\hat{\boldsymbol{\beta}}}$.
The null model includes only a bias term (intercept), and thus assumes an uninformative constant selection probability across all injected events. While this captures the global detection probability, the null model does not characterize the location or steepness of the transition region in parameter space, nor any curvature or interactions among predictors. Typical values of $R^2_{\text{McF}}$ for well-fit logistic models fall between 0.2 and 0.4, with values closer to 1 indicating a stronger fit \citep{McFadden_1979}. We report the calculated value for each model in Table \ref{tab:r2_brier_summary}.

\subsection{Pearson Residuals}
We also compute the Pearson residuals to assess the goodness-of-fit for each logistic regression model across different polynomial orders. For each injection $i$, the Pearson residual quantifies the deviation between the observed binary outcome $y_i$ and the predicted selection probability $S(\mathbf{x}_i)$. To obtain a single summary statistic for each polynomial order model we calculate the residual sum of squares (RSS), given by

\[
{\rm RSS} = \sum_{i=1}^N \frac{[y_i - S(\mathbf{x}_i)]^2}{S(\mathbf{x}_i)\left[ 1-S(\mathbf{x}_i)\right]},
\]
where the denominator corresponds to the standard deviation of a Bernoulli random variate with probability $S(\mathbf{x}_i)$. Unlike in linear regression, the logistic regression model does not assume homoskedastic gaussian errors across the model domain. 

\subsection{Brier Score}

We compute the Brier score as a metric for evaluating the generalized probabilistic accuracy of each fitted selection function. The Brier score is defined as the mean squared error between the predicted selection probabilities and the observed binary outcomes:
\[
\mathrm{BS} = \frac{1}{N} \sum_{i=1}^N \left( S(\mathbf{x}_i) - y_i \right)^2.
\]
This score ranges from 0 to 1, with lower values indicating better predictive performance. In the context of logistic regression, the Brier score can be interpreted as a proper scoring rule for probabilistic forecasts \citep{Brier1950}.

\bibliography{sample631}{}
\bibliographystyle{aasjournal}



\end{document}